\documentclass[fleqn,10pt]{wlscirep}
\usepackage[utf8]{inputenc}
\usepackage[T1]{fontenc}
\title{Thermally activated intermittent dynamics\\of creeping crack fronts along disordered interfaces}

\author[1,2,*]{Tom Vincent-Dospital}
\author[1,*]{Alain Cochard}
\author[3,4]{Stéphane Santucci}
\author[2]{Knut J\o rgen M\aa l\o y}
\author[1,2,*]{Renaud Toussaint}
\affil[1]{Université de Strasbourg, ITES UMR 7063, Strasbourg F-67084, France}
\affil[2]{SFF Porelab, The Njord Centre, Department of physics, University of Oslo,  Norway}
\affil[3]{Université de Lyon, ENS de Lyon, Université Claude Bernard, CNRS, Laboratoire de Physique, France}
\affil[4]{Lavrentyev Institute of Hydrodynamics, Siberian Branch of the Russian Academy of Sciences, Novosibirsk, Russia}
\affil[*]{tom.vincent-dospital@fys.uio.no, alain.cochard@unistra.fr and renaud.toussaint@unistra.fr}

\begin{abstract}
          ABSTRACT. We present a subcritical fracture growth model, coupled with the elastic redistribution of the acting mechanical stress along rugous rupture fronts. We show the ability of this model to quantitatively reproduce the intermittent dynamics of cracks propagating along weak disordered interfaces. To this end, we assume that the fracture energy of such interfaces (in the sense of a critical energy release rate) follows a spatially correlated normal distribution.
          We compare various statistical features from the obtained fracture dynamics to that from cracks propagating in sintered polymethylmethacrylate (PMMA) interfaces.
          In previous works, it has been demonstrated that such an approach could reproduce the mean advance of fractures and their local front velocity distribution. Here, we go further by showing that the proposed model also quantitatively accounts for the complex self-affine scaling morphology of crack fronts and their temporal evolution, for the spatial and temporal correlations of the local velocity fields and for the avalanches size distribution of the intermittent growth dynamics.
          We thus provide new evidence that Arrhenius-like subcritical growth laws are particularly suitable for the description of creeping cracks.
\end{abstract}

\usepackage{color} 
\usepackage{cancel} 
\usepackage[utf8]{inputenc} 
\usepackage{csquotes} 
\usepackage[caption = false]{subfig} 
\usepackage{hyperref} 
\hypersetup{colorlinks=true, linkcolor=blue, filecolor=magenta, urlcolor=blue}  
\usepackage{graphicx}
\usepackage{dcolumn}
\usepackage{bm}
\usepackage{textcomp}
\usepackage{upgreek}
\usepackage{float}

\usepackage[section]{placeins} 

\begin{document}

\flushbottom
\maketitle
\thispagestyle{empty}

\section{Introduction}

In the physics of rupture, understanding the effects that material disorder has on the propagation of cracks is of prime interest. For instance, the overall strength of large solids is believed to be ruled by the weakest locations in their structures, and notably by the voids in their bulk samples\cite{lawn_1993,porosity_crack_initiation}. There, cracks tend to initiate as the mechanical stress is concentrated.
A growing focus has been brought on models in which the description of the breaking matrix remains continuous (i.e., without pores). There, the material disorder resides in the heterogeneities of the matrix\cite{GaoRice_redistribution, Steph_PRL_crackling, Laurson2010, FBM_Schmittbuhl, Crackling_fiber_bundle, Cochard, Wiese2021}. The propagation of a crack is partly governed by its spatial distribution in surface fracture energy, that is, the heterogeneity of the energy needed to generate two opposing free unit surfaces in the continuous matrix\cite{Griffith1921}, including the dissipation processes at the tip\cite{Irwin1957}. From this disorder, one can model a rupture dynamics which holds a strongly intermittent behaviour, with extremely slow (i.e., pinned) and fast (i.e., avalanching) propagation phases. In many physical processes, including\cite{Bares14, crackling_crack2, crackling_crack1} but not limited\cite{crackling_imbibition, crackling_martensitic, crackling_plastic, crackling_Barkhausen} to the physics of fracture, such intermittency is referred to as crackling noise\cite{crackling_nat, avalanche_classes}. In the rupture framework, this crackling noise is notably studied to better understand the complex dynamics of geological faults\cite{crackling_fault_1, crackling_fault_2, crackling_fault_3, crackling_fault_4, crackling_fault_5}, and their related seismicity.\\
Over the last decades, numerous experiments have been run on the interfacial rupture of oven-sintered acrylic glass bodies (PMMA)\cite{Maloy2006, Santucci_2010, Tallakstad2011}. Random heterogeneities in the fracture energy were introduced by sand blasting the interface prior to the sintering process. An important aspect of such experiments concerns the samples preparation, which allows to constrain the crack to propagate along a weak (disordered) plane. It simplifies the fracture problem, leading to a negligible out-of plane motion of the crack front. This method has allowed to study the dynamics of rugous fronts, in particular because the transparent PMMA interface becomes more opaque when broken. Indeed, the generated rough air-PMMA interfaces reflect more light, and the growth of fronts can thus be monitored.\\
Different models have successfully described parts of the statistical features of the recorded crack propagation. Originally, continuous line models\cite{Tanguy1998,Steph_PRL_crackling,Laurson2010,avalanche_classes} were derived from linear elastic fracture mechanics. While they could reproduce the morphology of slow rugous cracks and the size distribution of their avalanches, they fail to account for their complete dynamics and, in particular, for the distribution of local propagation velocity and for the mean velocity of fronts under different loading conditions. Later on, fiber bundle models were introduced\cite{FBM_Schmittbuhl,a_numerical_study,bundle2}, where the fracture plane was discretized in elements that could rupture ahead of the main front line, allowing the crack to propagate by the nucleation and the percolation of damage. The local velocity distribution could then be reproduced, but not the long term mean dynamics of fronts at given loads. The most recent model (Cochard \textit{et al.}\cite{Cochard}) is a thermally activated model, based on an Arrhenius law, where the fracture energy is exceeded at subcritical stresses due to the molecular agitation. It contrasts to other models that are  strictly threshold based (the crack only advances when the stress reaches a local threshold, rather than its propagation being subcritical). A notable advantage of the subcritical framework is that its underlying processes are, physically, well understood, and Arrhenius-like laws have long shown to describe various features of slow fracturing processes\cite{Brenner, Zhurkov1984, Santucci2004, Santucci_2006, Maloy2006, Vanel_2009}.
In particular, this framework has proven to reproduce both the mean behaviour of experimental fronts\cite{Lengline2011} (i.e., the average front velocity under a given load) and the actual distributions of propagation velocities along these fronts\cite{Cochard}, whose fat-tail is preserved when observing cracks at different scales\cite{NonGaussian_Fracture}. It has recently been proposed\cite{TVD1,TVD2} that the same model might also explain the faster failure of brittle matter, that is, the dramatic propagation of cracks at velocities close to that of mechanical waves, when taking into account the energy dissipated as heat around a progressing crack tip. Indeed, if fronts creep fast enough, their local rise in temperature becomes significant compared to the background one, so that they can avalanche to a very fast phase, in a positive feedback loop\cite{TVD1,TVD2}.\\
Here, we only consider slow fronts (i.e., fronts that creep slowly enough so that their temperature elevation is supposed to remain negligible). Building on the work of Cochard \textit{et al.}\cite{Cochard}, we study various statistical features that can be generated with this Arrhenius-based model (re-introduced in section\,\ref{sec:model}), when simulating the rupture of a disordered interface. By comparing these features to those reported for the PMMA experiment by Tallakstad \textit{et al.}\cite{Tallakstad2011,NonGaussian_Fracture}, Santucci \textit{et al.}\cite{Santucci_2010} and Maløy \textit{et al.}\cite{Maloy2006}, we show a strong match to the experimental data for many of the scaling laws describing the fracture intermittent dynamics, including the growth of the fracture width (section\,\ref{sec:growth}), its distribution in local propagation velocity (section\,\ref{sec:vel}), the correlation of this velocity in space and time (section\,\ref{sec:cor}), the size of the propagation avalanches (section\,\ref{sec:aval}) and the front Hurst exponents (section\,\ref{sec:hurst}). We hence re-enforce the relevance of simple thermodynamics coupled with elasticity in the description of material failure.

\section{Propagation model\label{sec:model}}

\subsection{Constitutive equations}

\begin{figure}
\centering
\includegraphics[width=1\linewidth]{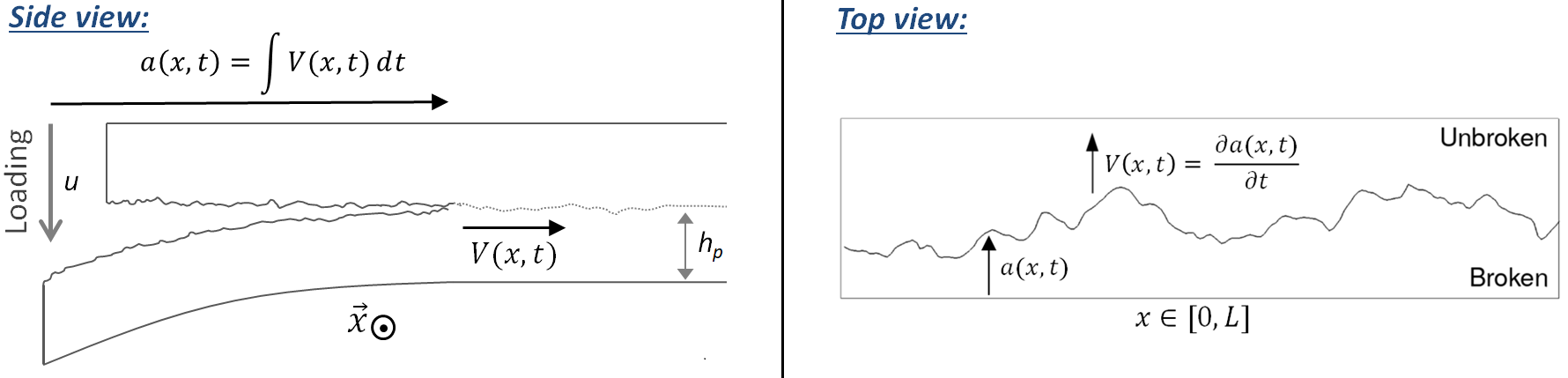}
  \caption{(Left): Separation of two rugous and sintered PMMA plates, as reported by Tallakstad \textit{et al.}\cite{Tallakstad2011}. The rugosity of the (quasi-plane) interface is here massively exaggerated (the plates are centimetres thick while the standard deviation in the interface topography is less than a micrometer\cite{interf_imaging}). A local position of the front has an advancement $a(x,t)$ and a velocity $V(x,t)$. The out of frame coordinate is $x$ and $t$ is the time variable. (Right): top view, showing the crack font roughness, which arises from the disorder in the interface's fracture energy.
  }
 \label{pmma_sep}
\end{figure}

We consider rugous crack fronts that are characterised by a varying and heterogeneous advancement $a(x,t)$ along their front, $x$ being the coordinate perpendicular to the average crack propagation direction, $a$ the coordinate along it, and $t$ being the time variable (see Fig.\,\ref{pmma_sep}). At a given time, the velocity profile along the rugous front is modelled to be dictated by an Arrhenius-like growth, as proposed by Cochard \textit{et al.}\cite{Cochard}:
\begin{equation}
   V(x,t) = V_0 \text{ min}\left[ \exp \left( {-\cfrac{\alpha^2[G_c(x,a)-G(x,t)]}{k_B T_0}} \right), 1\right],
   \label{model1}
\end{equation}
where $V(x,t)=\partial a(x,t)/\partial t$ is the local propagation velocity of the front at a given time and $V_0$ is a nominal velocity, related to the atomic collision frequency\cite{kinetics}, which is typically similar to the Rayleigh wave velocity of the medium in which the crack propagates\cite{Freund1972}. The exponential term is a subcritical rupture probability (i.e., between $0$ and $1$). It is the probability for the rupture activation energy (i.e., the numerator term in the exponential) to be exceeded by the thermal bath energy $k_B T_0$, that is following a Boltzmann distribution\cite{kinetics}. The Boltzmann constant is denoted $k_B$ and the crack temperature is denoted $T_0$ and is modelled to be equal to a constant room temperature (typically, $T_0=298\,$K). 
It corresponds to the hypothesis that the crack is propagating slowly enough so that no significant thermal elevation occurs by Joule heating at its tip (i.e., as inferred by Ref.\cite{TVD1, TVD2}). Such propagation without significant heating is notably believed to take place in the experiments by Tallakstad \textit{et al.}\cite{Tallakstad2011} that we here try to numerically reproduce, and whose geometry is shown in Fig.\,\ref{pmma_sep}. Indeed, their reported local propagation velocities $V$ did not exceed a few millimetres per second, whereas a significant heating in acrylic glass is only believed to arise for fractures faster than a few centimetres per second\cite{TVD2,PMA3ss}. See the supplementary information for further discussion on the temperature elevation.\\
In Eq.\,(\ref{model1}), the rupture activation energy is proportional to the difference between an intrinsic material surface fracture Energy $G_c$ (in J\,m\textsuperscript{-2}) and the energy release rate $G$ at which the crack is mechanically loaded, which corresponds to the amount of energy that the crack dissipates to progress by a given fracture area. The front growth being considered subcritical, we have $G<G_c$. We here model the fracture energy $G_c$ to hold some quenched disorder that is the root cause for any propagating crack front to be rugous and to display an intermittent avalanche dynamics. This disorder is hence dependent on two position variables along the rupture interface. For instance, at a given front advancement $a(x,t)$, one gets $G_c=G_c(x,a)$. The coefficient $\alpha^2$ is an area which relates the macroscopic $G$ and $G_c$ values to, respectively, the microscopic elastic energy $U=\alpha^2 G$ stored in the molecular bonds about to break, and to the critical energy $U_c=\alpha^2 G_c$ above which they actually break. See Vanel \textit{et al.}\cite{Vanel_2009}, Vincent-Dospital \textit{et al.}\cite{TVD2} or the supplementary information for more insight on the $\alpha^2$ parameter, which is an area in the order of $d_0^3/l$, where $d_0$ is the typical intra-molecular distance and $l$ is the core length scale limiting the stress divergence at the crack tip.\\
Finally, the average mechanical load that is applied on the crack at a given time is redistributed along the evolving rugous front, so that $G=G(x,t)$. To model such a redistribution, we here use the Gao and Rice\cite{GaoRice_redistribution} formalism, which integrates the elastostatic kernel along the front:
\begin{equation}
   G(x,t) = \overline{G}(t) \left[1 - \frac{1}{\pi} \text{PV} \int_{-\infty}^{+\infty} \frac{\partial a(x',t) / \partial{x'}}{x-x'} \mathrm{d}x'\right].
   \label{model2}
\end{equation}
In this equation, $\overline{G}$ is the mean energy release rate and $\text{PV}$ stands for the integral principal value. We, in addition, considered the crack front as spatially periodic, which is convenient to numerically implement a spectral version of Eq.\,(\ref{model2}) \cite{Perrin_spectral_redistribution} as explained by Cochard \textit{et al.}\cite{Cochard}.\\
Equations (\ref{model1}) and (\ref{model2}) thus define a system of differential equations for the crack advancement $a$, which we have then solved with an adaptive time step Runge-Kutta algorithm\cite{DormandPrince}, as implemented by Hairer et al\cite{dop583}.

\subsection{Discretization\label{sec:discreti}}

\begin{figure}
\centering
\includegraphics[width=0.8\linewidth]{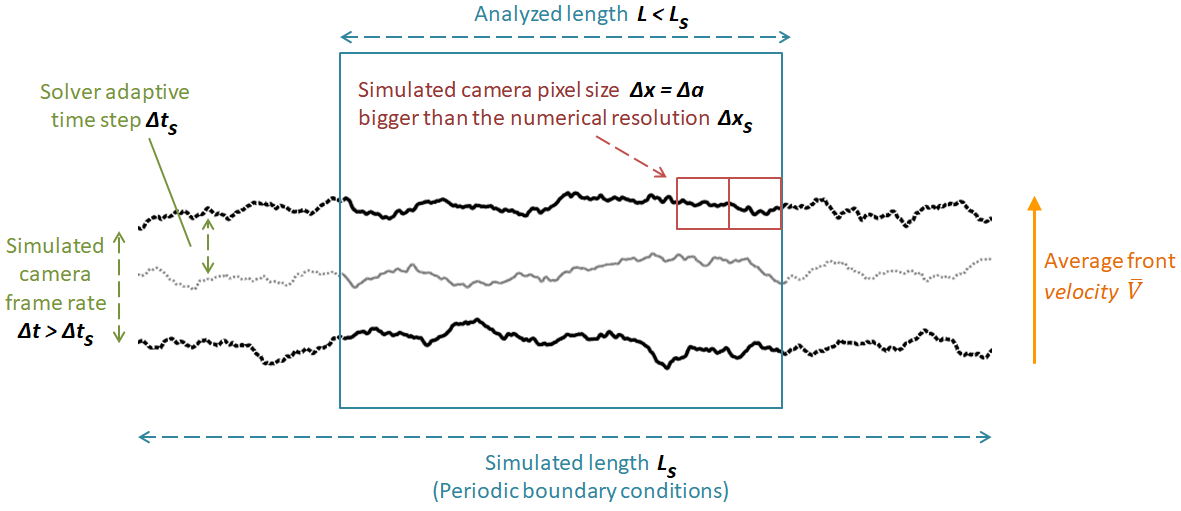}
  \caption{Illustration of the discretization principles and of the solver and observation grids. Three crack fronts at three successive times are shown, over which the parameters discussed in section\,\ref{sec:discreti} are defined.}
 \label{fig:discreti}
\end{figure}
In this section, we discuss the main principles we have used in choosing the numerical accuracy of our solver. The related parameters are illustrated in Fig.\,\ref{fig:discreti}.\\
In attempting to correctly reproduce the experimental results of Tallakstad \textit{et al.}\cite{Tallakstad2011}, this solver needs to use space and time steps, here denoted $\Delta x_s$ and $\Delta t_s$, at least smaller than those on which the experimental fronts were observed and analysed. Thus, $\Delta x_s$ needs to be smaller than the experimental resolutions in space (the camera pixel size) $\Delta x=\Delta a$ of about $2$ to $5\,\upmu$m and $1/\Delta t_s$ needs to be higher than the experimental camera frame rate $1/\Delta t$. This frame rate was set by Tallakstad \textit{et al.}\cite{Tallakstad2011} to about $(100\overline{V})/\Delta x$, where $\overline{V}$ is the average front velocity of a given fracture realisation.
The propagation statistics of our simulated fronts, henceforward shown in this manuscript, have, for consistency, always been computed on scales comparable to the experimental $\Delta x$, $\Delta a$, $\Delta t$ steps. Thus, as $\Delta x_s<\Delta x$ and $\Delta t_s<\Delta t$, we have first decimated the dense numerical outputs on the experimental observation grid, by discarding smaller time scales and by averaging smaller space scales to simulate the camera frame rate and pixel size.\\
As the camera resolution was $1024$ pixels, the lengths $L$ of the crack segments that Tallakstad \textit{et al.}\cite{Tallakstad2011} analysed were $1024\Delta_x=3$ to $7$\,mm long, and we have then analysed our numerical simulations on similar front widths. Yet, these simulations were priorly run on longer front segments, $L_s>L$, in order to avoid any possible edge effects in the simulated crack dynamics (for instance in the case where $L$ would not be much bigger than the typical size of the $G_c$ quenched disorder).\\
Overall, we have checked that the numerical results presented henceforward were obtained using a high enough time and space accuracy for them to be independent of the discretization (see the supplementary information).

\subsection{Physical parameters values}

For the model dynamics to be compared to the experiments\cite{Tallakstad2011}, one must also ensure that the $V_0$, $\alpha$, $T_0$, $G$ and $G_c$ parameters are in likely orders of magnitude.\\
As $V_0$ is to be comparable to the Rayleigh velocity of acrylic glass, we have here used $1$\,km\,s\textsuperscript{-1}\cite{plexi}. Lengliné \textit{et al.}\cite{Lengline2011} furthermore estimated the ratio $\alpha^2/(k_B T_0)$ to be about $0.15$\,m\textsuperscript{2}\,J\textsuperscript{-1} and they could approximate the quantity $V_0\exp(-\alpha^2 \overline{G_c}/[k_B T_0])$ to about $5\times10^{-14}$\,m\,s\textsuperscript{-1}, where $\overline{G_c}$ is the average value of $G_c$. With our choice on the value of $V_0$, we then deduce $\overline{G_c}\sim250$\,J\,m\textsuperscript{-2} (Note that the trade-off between $V_0$ and $G_c$ should be kept in mind when comparing our results with those by Cochard \textit{et al.}\cite{Cochard}, as both papers use a different $V_0$.)
The value thus inverted for the fracture energy ($250$\,J\,m\textsuperscript{-2}), that is to represent the sintered PMMA interfaces, is logically smaller but comparable to that inferred by Vincent-Dospital \textit{et al.}\cite{TVD2} for the rupture of bulk PMMA (about $1300$\,J\,m\textsuperscript{-2}). Qualitatively, the longer the sintering time, the closer one should get from such a bulk toughness, but the less likely an interfacial rupture will occur.\\
Experiments in two different regimes were run\cite{Tallakstad2011}: a forced one where the deflection of the lower plate (see Fig.\,\ref{pmma_sep}) was driven at a constant speed, and a relaxation regime, where the deflection was maintained constant while the crack still advances. In both scenarii, the long term evolution of the average load $\overline{G}(t)$ and front position $\overline{a}(t)$ was shown\cite{Lengline2011,Cochard} to be reproduced by Eq.\,(\ref{model1}). In the case of the experiments of Tallakstad \textit{et al.}\cite{Tallakstad2011}, the intermittent dynamics measured in the two loading regimes were virtually identical. Such similarity likely arises from the fact that $\overline{G}$ was, in both cases, computed to be almost constant over time, in regard to the spatial variation in $G$, described by Eq.\,(\ref{model2}) (see the supplementary information). Here, we will then consider that the crack is, in average along the front, always loaded with the same intensity (i.e., $\overline{G}(t)=\overline{G}$).\\
The actual value of $\overline{G}$, together with the average surface fracture energy of the medium $\overline{G_c}$, then mainly controls the average crack velocity $\overline{V}$. This average velocity was investigated over five orders of magnitude in Ref.\cite{Tallakstad2011}, from $0.03$ to $140\,\upmu$m\,s\textsuperscript{-1}, which, in our formalism, shall correspond to values of $(\overline{G_c}-\overline{G}$) between $145$ and $85$\,J\,m\textsuperscript{-2}, respectively, which is consistent with the values of $\overline{G}$ measured by Lengliné \textit{et al.}\cite{Lengline2011} for cracks propagating at similar speeds. The intermittency of the crack motion was experimentally inferred to be independent on $\overline{V}$ and we show, in the supplementary information, that it is also the case in our simulations. The velocity variation along the front shall then only arise from the disorder in $G_c$ and from the related variations of $G$ due to the roughness of the crack front. Further in this manuscript, we will use $\overline{G}=120$\,J\,m\textsuperscript{-2}, which corresponds to an average propagation velocity of about $1.5\,\upmu$m\,s\textsuperscript{-1}.

\section{Heterogeneous fracture energy}

\begin{figure}
\centering
\includegraphics[width=1\linewidth]{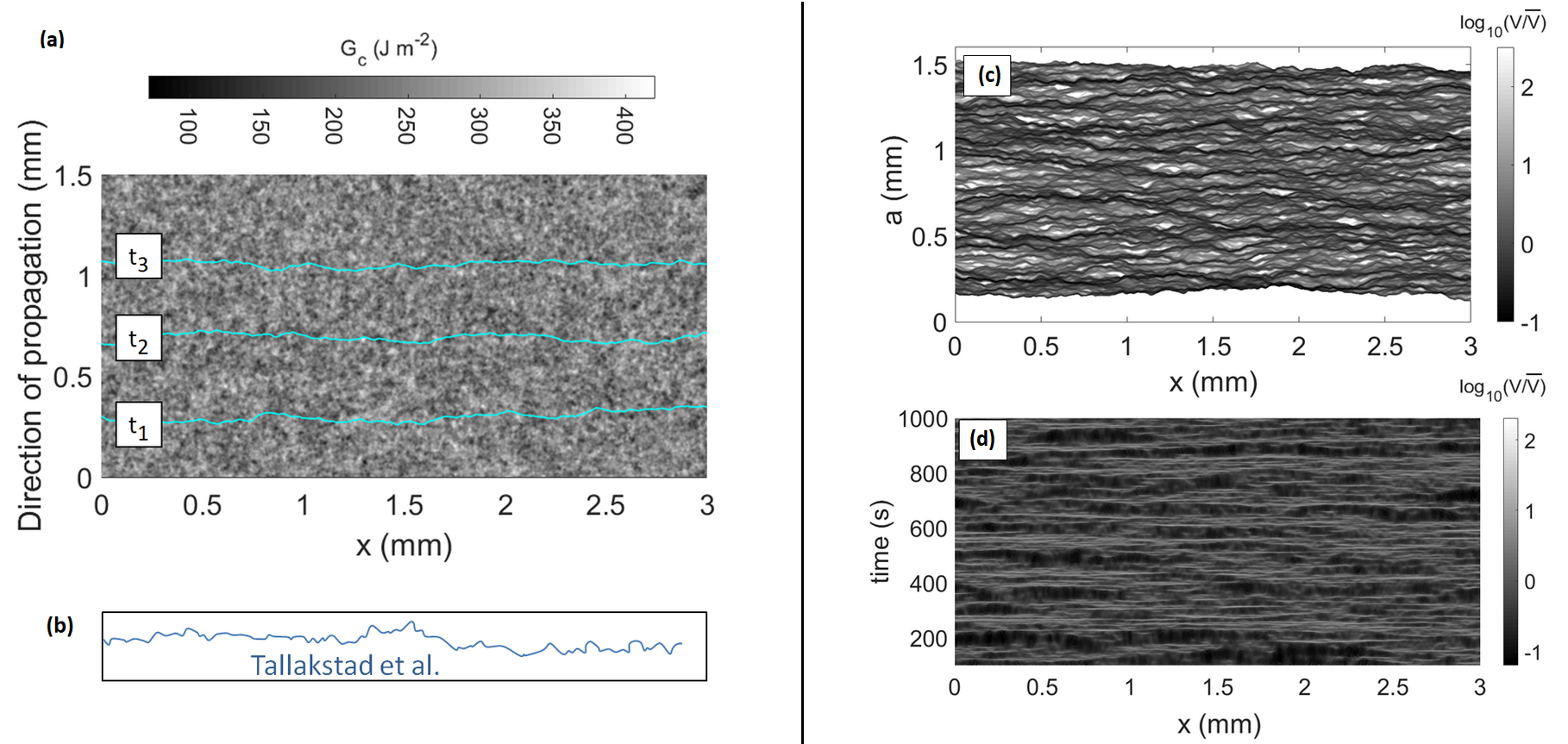}
  \caption{(a): Normal distribution of the fracture energy $G_c$ considered for the simulations. The average value is $\overline{G_c}=250$\,J\,m\textsuperscript{-2}, with a standard deviation $\delta G_c=35$\,J\,m\textsuperscript{-2} and a correlation length $l_c=50\,\upmu$m. The three lines are the modelled propagating front at three different times $t_1<t_2<t_3$, using Eqs.\,(\ref{model1}) and\,(\ref{model2}). (b): A crack front reported by Tallakstad \textit{et al.}\cite{Tallakstad2011} (Fig. 3 of the experimental paper), plotted on the same spatial scales. (c and d): Local velocity maps $V(x,a)$ in the space-space domain (c) and $V(x,t)$ in the space-time domain (d) for a modelled crack propagating in this $G_c$ landscape. Both maps are shown with the same color scale and they are computed on a resolution similar to that of the experiments by Tallakstad \textit{et al.}\cite{Tallakstad2011}, using the waiting time matrix. The velocity are plotted related to the average crack velocity $\overline{V}=1.5\,\upmu$m\,s\textsuperscript{-1}. All parameters used to run the corresponding simulation are summarised in table\,\ref{tab_param}.}
 \label{fig:gc}
\end{figure}

Of course, the actual surface fracture energy field in which the rupture takes place will significantly impact the avalanches dynamics and the crack morphology. Such a field is yet a notable unknown in the experimental set-up of Tallakstad \textit{et al.}\cite{Tallakstad2011}, as their interface final strength derived from complex sand blasting and sintering processes. Although these processes were well controlled, so that the rough rupture experiments were repeatable, and although the surfaces prior to sintering could be imaged\cite{interf_imaging}, the actual resulting distribution in the interface cohesion was not directly measurable. While this is, to some extent, a difficulty in assessing the validity of the model we present, we will here show that a simple statistical definition of $G_c$ is enough to simulate most of the avalanches statistics.\\
We will indeed consider a normally distributed $G_c$ field around the mean value $\overline{G_c}$ with a standard deviation $\delta G_c$ and a correlation length $l_c$.
Such a landscape in $G_c$ is shown in Fig.\,\ref{fig:gc}a, and we proceed to discuss the chosen values of $\delta G_c$ and $l_c$ in sections\,\ref{sec:growth} and\,\ref{sec:vel}.

\subsection{Growth exponent and fracture energy correlation length\label{sec:growth}}

Among the various statistical features studied by Tallakstad \textit{et al.}\cite{Tallakstad2011}, was notably quantified the temporal evolution of their fracture fronts morphology. It was interestingly inferred that the standard deviation of the width evolution of a crack front $h$ scales with the crack mean advancement:
\begin{equation}
   \text{rms}\big(h(t)\big) = \sqrt{{<}{h(t)}^2{>}_{x,t_0}} \propto \big(\overline{V}t\big)^{\beta_G}.
   \label{growth}
\end{equation}
In this equation, $x$ is a given position along the front, $t$ is a time delay from a given reference time $t_0$, and $h$ writes as
\begin{equation}
   h_{x,t_0}(t) = \big[a(x,t_0+t)-\overline{a}(t_0+t)\big]-\big[a(x,t_0)-\overline{a}(t_0)\big],
   \label{growth2}
\end{equation}
$\overline{a}$ being the average crack advancement at a given time.
To mitigate the effect of the limited resolution of the experiments and obtain a better characterization of the scaling of the interfacial fluctuations on the shorter times, we computed the subtracted width,
\begin{equation}
   W(t) = \sqrt{\text{rms}(h(t))^2-\text{min}(\text{rms}(h(t)))^2},
   \label{growth3}
\end{equation}
as proposed in Barabasi and Stanley\cite{fractal_surface}, and done by Tallakstad \textit{et al.}\cite{Tallakstad2011} and Soriano \textit{et al.}\cite{cor_low_t}\\
\begin{figure}
\centering
\includegraphics[width=1\linewidth]{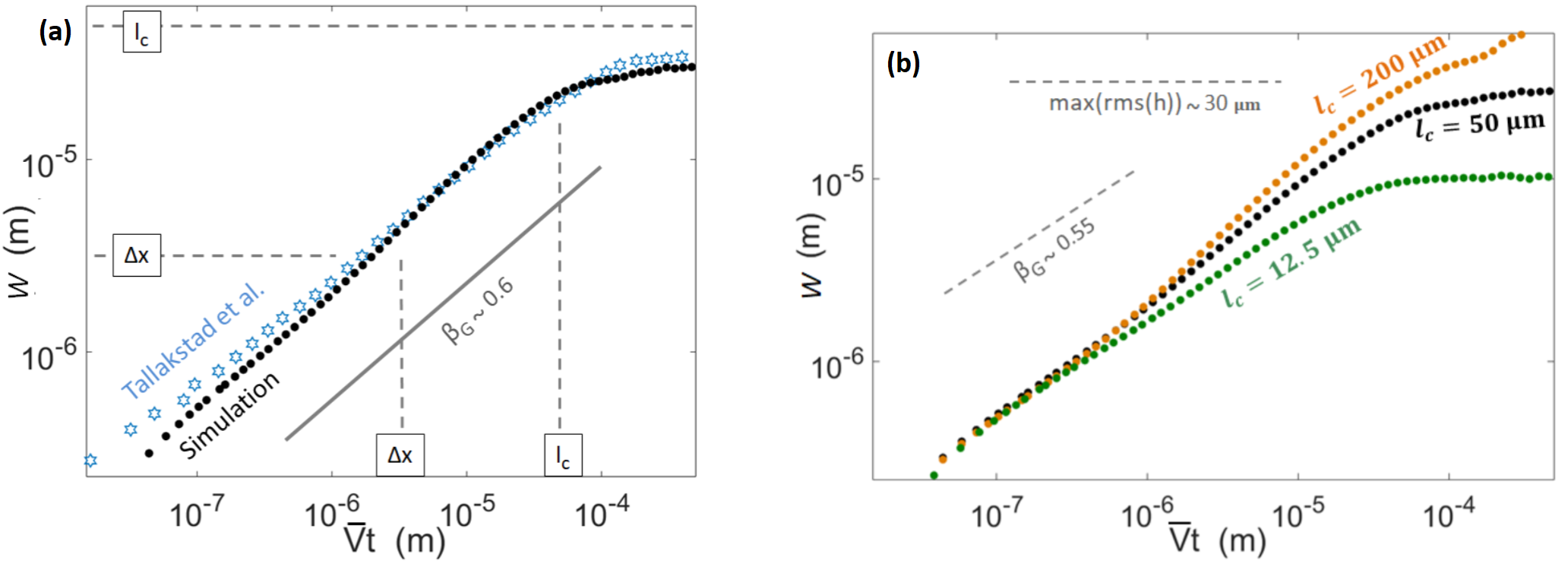}
  \caption{(a): Standard deviation of the width evolution of the crack front as a function of the mean crack advancement, as defined by Eqs.\,(\ref{growth}) to\,(\ref{growth3}) for the chosen simulation (plain points) and for the experiments\cite{Tallakstad2011} (hollow stars) (out of Fig.\,8, Expt.\,5 of the experimental paper). The continuous line has a slope $0.6$, close to that of the experimental points: $\beta_G\sim0.55$. The numerical $\beta_G$, obtained with a linear root mean square fit of the growth of $W$, is estimated as $\beta_G=0.60\pm0.05$. The dashed lines mark the observation scale $\Delta x$, corresponding to the experimental camera pixel size, and the chosen correlation length for the simulation $l_c=50\,\upmu$m. (b): The same width function for simulations with different correlation lengths $l_c$. The rest of the parameters are as defined in table\,\ref{tab_param}. The slope and plateau of the experimental data (shown in (a)) is marked by the dashed line for comparison.}
 \label{fig:growth}
\end{figure}
\begin{figure}
\centering
\includegraphics[width=1\linewidth]{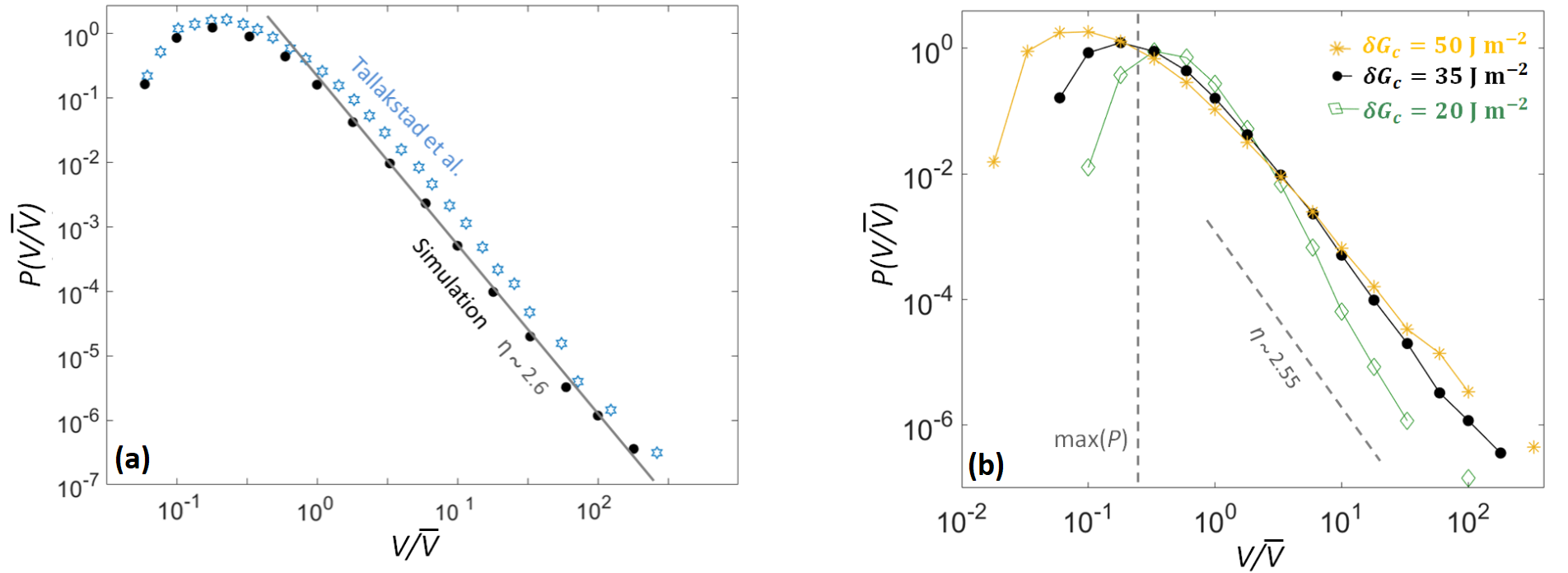}
  \caption{(a): Probability density function of the local propagation velocity along a simulated front (plain points), computed from the space-time map of Fig.\,\ref{fig:gc}d. The experimental probability\cite{Tallakstad2011} (out of Fig.\,5, Expt.\,5 of the experimental paper) is shown for comparison (hollow stars). The continuous line has a slope $-2.6$ close to that of the experimental points. This was achieved by setting the standard deviation for the disorder in fracture energy to $35$\,J\,m\textsuperscript{-2}. The numerical $\eta$, obtained with a linear root mean square fit of the distribution tail, is estimated as $\eta=2.6\pm0.1$. (b): The same distribution for three three simulations with different values of $\delta G_c$. We chose the value of $\delta G_c$ by tuning it and fitting the slope and maximum of the experimental data, which are illustrated by the dashed lines. The rest of the parameters used in these simulations are as defined in table\,\ref{tab_param}. Note that the ability of the model to reproduce the local velocity distribution was already shown by Cochard \textit{et al.}\cite{Cochard} (see text for explanation and discussion).}
 \label{fig:veldist}
\end{figure}
The scaling exponent $\beta_G$ is referred to as the growth exponent, and we will here show how it allows to deduce a typical correlation length for the interface disorder. 
Indeed, $\beta_G$ was measured to be $0.55\pm0.08$ by Tallakstad \textit{et al.}\cite{Tallakstad2011}. This value is close to\,$1/2$, that is, consistent with an uncorrelated growth process (e.g.,\cite{fractal_surface}), such as simple diffusion or Brownian motion. We thus get a first indication on the disorder correlation length scale $l_c$. To display an uncorrelated growth when observed with the experimental resolution ($\Delta x\sim 3\,\upmu$m), the fronts likely encountered asperities whose size was somewhat comparable to this resolution. Indeed, if these asperities in $G_c$ were much bigger, the growth would be perceived as correlated. By contrast, if they were much smaller (orders of magnitude smaller), the rugosity of the front would not be measurable, as only the average $\overline{G_c}$ over an observation pixel would then be felt. Furthermore, and as shown in Fig.\,\ref{fig:growth}a, the exponent $\beta_G$ was observed on scales $(\overline{V}t)$ up to $100\,\upmu$m, above which $W$ stabilised to a plateau value of about $30\,\upmu$m. A common picture is here drawn, as both this plateau value and the typical crack propagation distance at which it is reached are likely to be correlated with $l_c$, as the front is to get pinned on the strongest asperities at this scale.\\
From all these clues, we have considered, in our simulations, the correlation length of the disorder to be about $l_c=50\,\upmu$m, and we show in Fig.\,\ref{fig:growth}a that it allows an approximate reproduction of the front growth exponent and of the plateau at high $\overline{V}t$. The accuracy reported for the exponents in this manuscript is estimated by fitting various portions of the almost linear data points and reporting the dispersion of the thus inverted slopes. In Fig.\,\ref{fig:growth}b, we also show how varying $l_c$ impacts $W$, and, in practice, we have chosen $l_c$ by tuning it when comparing these curves to the experimental one. Noteworthily, the thus chosen $l_c$ is in the lower range of the size of the blasting sand grains ($50-300\,\upmu$m) that were used\cite{Tallakstad2011} to generate the interface disorder. It is also comparable to the correlation length of the blasting induced topographic anomalies $\sim18\,\upmu$m on the post-blasting/pre-sintering PMMA surfaces, as measured by Santucci \textit{et al.}\cite{interf_imaging} by white light interferometry.

\begin{table}
\centering
\begin{tabular}{c c c c}
&\textbf{\,Parameter\,} & \textbf{\,\,\,\,\,\,\,\,\,Value\,\,\,\,\,\,\,\,\,} & \textbf{\,\,\,\,\,\,\,\,\,Unit\,\,\,\,\,\,\,\,\,} \\ \hline 
&$V_0$               & $1000$ & m s\textsuperscript{-1}        \\ 
&$\alpha^2/(k_B T_0)$       & $0.15$ & m\textsuperscript{2}\,J\textsuperscript{-1}  \\ 
&$\overline{G_c}$    & $250$    & J m\textsuperscript{-2}         \\ 
\textbf{(a)}\,\,\,\,\,\,\,\,\,&$\overline{G}$      & $120$    & J m\textsuperscript{-2}         \\
&$\delta G_c$      & $35$    & J m\textsuperscript{-2}         \\ 
&$l_c$      & $50$    & $\upmu$m         \\  \hline
&$\Delta a = \Delta x$     & $3$  &  $\upmu$m     \\
\textbf{(b)}\,\,\,\,\,\,\,\,\,&$\Delta t$     & $10$  &  ms     \\
&$L$     & $3000$  &  $\upmu$m     \\ \hline
&$\Delta x_s$     & $1$  &  $\upmu$m     \\
\textbf{(c)}\,\,\,\,\,\,\,\,\,&$\Delta t_s$     & $\sim5$   &  ms     \\
&$L_s$     & $6000$  &  $\upmu$m     \\
\end{tabular}
\caption{Summary of all parameters that are considered in this manuscript. (a): physical parameters in Eqs.\,(\ref{model1}) and\,(\ref{model2}) believed to be representative of the studied creep experiments. (b): observation scale of the modelled fronts, similar to the experimental ones of Tallakstad \textit{et al.}\cite{Tallakstad2011}. (c): the solver grid, finer than the observation scale for numerical accuracy.}
\label{tab_param}
\end{table}

\subsection{Local velocity distribution and fracture energy standard deviation\label{sec:vel}}

While the crack advances at an average velocity $\overline{V}$, the local velocities along the front, described by Eq.\,(\ref{model1}), are, naturally, highly dependent on the material disorder: the more diverse the met values of $G_c$ the more distributed shall these velocities be.\\
Maløy \textit{et al.}\cite{Maloy2006} and Tallakstad \textit{et al.}\cite{Tallakstad2011} inferred the local velocities of their cracks with the use of a so-called waiting time matrix. That is, they counted the number of discrete time steps a crack would stay on a given camera pixel before advancing to the next one. They then deduced an average velocity for this pixel by inverting this number and multiplying it by the ratio between the pixel size and the time between two pictures: $\Delta a/\Delta t$. Such a method, that provides a spatial map $V(x,a)$, was applied to our simulated fronts, and we show this $V(x,a)$ map in Fig.\,\ref{fig:gc}c. As to any time $t$ corresponds a front advancement $a(x,t)$ (recorded with a resolution $\Delta a$), an equivalent space-time map $V(x,t)$ can also be computed, and it is shown in Fig.\,\ref{fig:gc}b. The experimental report\cite{Tallakstad2011} presented the probability density function of this latter (space-time) map $P(V/\overline{V})$, and it was inferred that, for high values of $V$, the velocity distribution scaled with a particular exponent $\eta = 2.6 \pm 0.15$\cite{Tallakstad2011,NonGaussian_Fracture} (see Fig.\,\ref{fig:veldist}a). That is, it was observed that
\begin{equation}
   P\left({V}/{\overline{V}}\right) \propto \left({V}/{\overline{V}}\right)^{-\eta}.
   \label{vel_tail}
\end{equation}
Cochard \textit{et al.}\cite{Cochard}, who introduced the model that we here discuss, inferred that the $\eta$ exponent was mainly depending on $\alpha^2(\delta G_c)^2/[k_B T_0(\overline{G_c})^2]$. Truly, a more comprehensive expression could also include other quantities, such as $V_0$ or $l_c$. Yet, as all other parameters have now been estimated, we can deduce $\delta G_c$ by varying it to obtain $\eta\sim2.6$. We show, in Fig.\,\ref{fig:veldist}b, how varying $\delta G_c$ impacts $P(V/\overline{V})$ and $\eta$. We found $\delta G_c\sim35$\,J\,m\textsuperscript{-2}. In Fig.\,\ref{fig:veldist}a, we show the corresponding velocity distribution for a simulation run with this parameter, together with that from Tallakstad \textit{et al.}\cite{Tallakstad2011}, showing a good match. Note that the ability of the model to reproduce the local velocity distribution was already shown by Cochard \textit{et al.}\cite{Cochard}, and this figure mainly aims at illustrating our calibration of the fracture energy field. The model we present is also slightly different to that of Cochard et al., as the interface fracture energy is now described at scales below its correlation length, similarly to the observation scale of the experiments. We here verify that the reproduction of the local velocity distribution is still valid at these small scales. Satisfyingly, the inverted value of $\delta G_c$ is not too far from the value found by Lengliné \textit{et al.}\cite{Lengline2011} for their fluctuation in the mean fracture energy $\overline{G_c}$ along their sintered plates, when studying the mean front advancement (i.e., neglecting the crack rugosity) in similar PMMA interfaces, which was about $25$\,J\,m\textsuperscript{-2}.

\section{Further statistics}

We have now estimated the orders of magnitude of all parameters in Eqs.\,(\ref{model1}) and\,(\ref{model2}), including a likely distribution for an interface fracture energy representative of the experiments\cite{Tallakstad2011} we aim to simulate (i.e., $\overline{G_c}$, $\delta G_c$ and $l_c$). For convenience, this information is summarised in table \ref{tab_param}.\\
We will now pursue by computing additional statistics of the crack dynamics to compare them to those reported by Tallakstad \textit{et al.}\cite{Tallakstad2011}.

\subsection{Local velocities correlations\label{sec:cor}}

\begin{figure}
\includegraphics[width=1\linewidth]{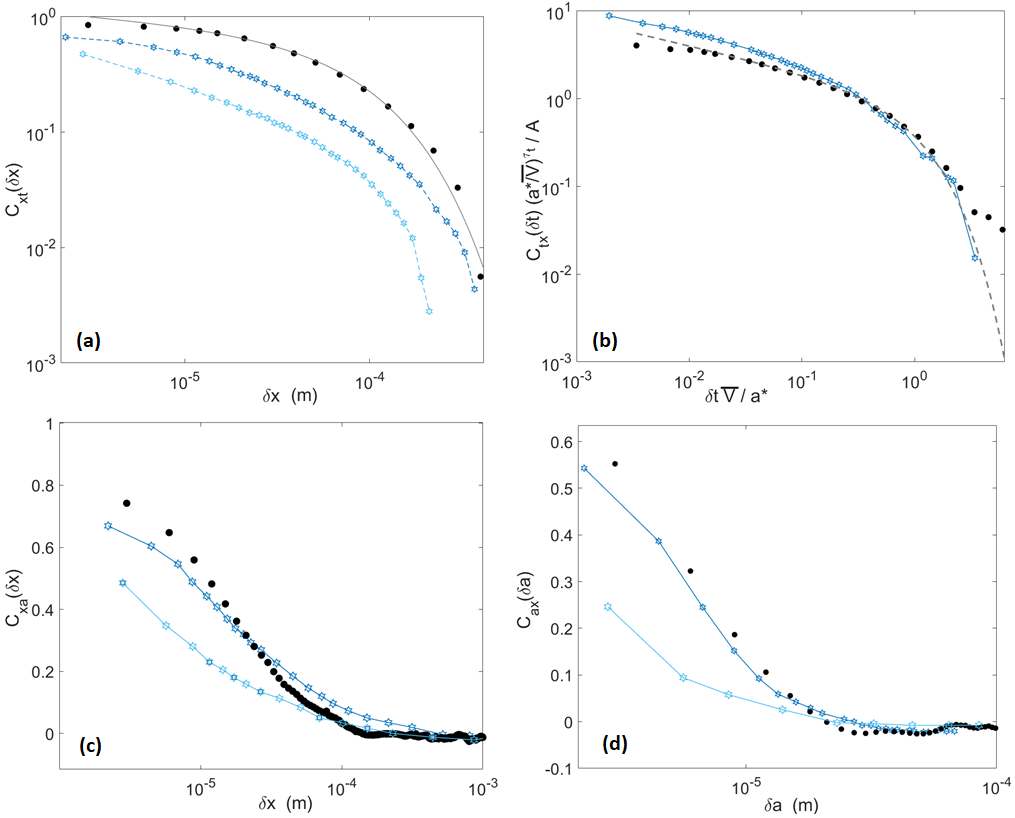}
  \caption{Local velocity correlation functions in space and time as defined by Eq.\,(\ref{correlation2}). The plain points were computed from the simulation whose parameters are presented in table\,\ref{tab_param} and the hollow stars are some of the experimental data points extracted from Figs.\,6 and\,7 of Tallakstad \textit{et al.}\cite{Tallakstad2011}. In plot (a), the line overlying the numerical data set corresponds to a fit using Eq.\,(\ref{CxtMod}). (b) is plotted in a domain that allowed a good collapse of the experimental data for many experiments\cite{Tallakstad2011}. The parameters $A$, $a^*$ and $\tau_t$ were inverted from Eq.\,(\ref{CtMod}), and the related fit is shown by the dashed line overlying the numerical data set. Plots (a), (c) and (d) hold two curves for the experiments, corresponding to two distinct sets of experiments done on two different sintered PMMA bodies. (b) shows Expt.\,5 of Tallakstad \textit{et al.}\cite{Tallakstad2011}.}
 \label{fig:cor}
\end{figure}
In particular, we here compute the space and time correlations of the velocities along the front. That is, four correlation functions that are calculated from the  $V(x,t)$ and $V(x,a)$ matrices (shown in Fig.\,\ref{fig:gc}) and defined as:
\begin{equation}
   C_{ij}(\delta i)=\left\langle\frac{\left[V(i_0+\delta i,j)-\overline{V_j}\,\right]\left[V(i_0,j)-\overline{V_j}\,\right]}{{(\delta V_j)}^2}\right\rangle_{i_0},
   \label{correlation2}
\end{equation}
where $i$ and $j$ are the variables of either $V(x,t)$ or $V(x,a)$ and $\delta i$ a given $i$ increment. $\overline{V_j}$ is the mean of $V(i, j)$ taken along $j$ at a given $i_0$. The corresponding $\delta V_j$ is the velocity standard deviation along the same direction and for the same $i_0$. The correlation functions hence defined are the same as those used by Tallakstad \textit{et al.}\cite{Tallakstad2011} on their own data, allowing to display a direct comparison of them in Fig.\,\ref{fig:cor}. A good general match is obtained.\\
One can notice the comparable cut-offs along the $x$ axis (Fig.\,\ref{fig:cor}\,a and\,c), indicating that our chosen correlation length for the interface disorder ($l_c$ inferred in section\,\ref{sec:growth}) is a good account of the experiment. Yet, one can notice that $C_{xt}$ (the velocity correlation along the crack front shown in Fig.\,\ref{fig:cor}\,a) is higher in the numerical case than in the experimental one. It could translate the fact that the experimental disorder holds wavelengths that are smaller than the observation scale $\Delta x$, and that our modelled $G_c$ distribution, where $l_c>\Delta x$, is rather simplified.\\
To go further, Tallakstad \textit{et al.}\cite{Tallakstad2011} modelled $C_{xt}$ as
\begin{equation}
   C_{xt}(\delta x)\propto{\delta x}^{-\tau_x}\exp\left(-\frac{\delta x}{x^*}\right),
   \label{CxtMod}
\end{equation}
and inverted the values of $\tau_x$ and $x^*$ to, respectively, $0.53$ and about $100\,\upmu$m. Doing a similar fit on the simulated data, we found $\tau_x\sim0.13$ and $x^*\sim94\,\upmu$m. The related function is displayed in Fig.\,\ref{fig:cor}\,a (plain line). Our small $\tau_x\sim0.13$ may derive, as discussed, from the better correlation that our simulation displays at small $\delta x$ ($\tau_x$ may in reality tend to zero for scales smaller than those we observe) compared to that of the experiments, while the matching $x^*$ probably relates to a satisfying choice we made for $l_c$. Overall, the existence of a clear scaling law at small offsets, as defined by Eq.\,(\ref{CxtMod}), is rather uncertain (see the two experimental plots in Fig.\,\ref{fig:cor}\,a) so that mainly the cut-off scale is of interest.\\
On the time correlation $C_{tx}$ (Fig.\,\ref{fig:cor}\,b), one can similarly define the parameters $A$, $\tau_t$ and $a^*$ to fit Eq.\,(\ref{correlation2}) with a function
\begin{equation}
   C_{tx}(\delta t)\approx A{\delta t}^{-\tau_t}\exp\left(-\frac{\overline{V}\delta t}{a^*}\right),
   \label{CtMod}
\end{equation}
where $A$ is a constant of proportionality. Fitting this function to Eq.\,(\ref{correlation2}) with a least-squares method, we found $\tau_t\sim0.3$ and $a^*\sim4.3\,\upmu$m, and this fit is represented by the dashed line in Fig.\,\ref{fig:cor}\,b. Tallakstad \textit{et al.}\cite{Tallakstad2011} reported $\tau_t\sim0.43$ and $a^*\sim7\,\upmu$m.
Figure\,\ref{fig:cor}\,b shows the experimental and simulated correlation functions in the $\overline{V}\delta t/a^*$ -- $C_{tx}(a^*/\overline{V})^{\tau_t}/A$ domain, as this allowed a good collapse of the data from numerous experiments\cite{Tallakstad2011}. We show that it also allows an approximate collapse of our modelled correlation on a same trend. Finally, the derived value of $a^*$ consistently matches the apparent cut-off length in the $C_ax$ correlation function in Fig.\,\ref{fig:cor}\,d. This length being of a magnitude similar to that of the observation scale $\Delta a$, the crack local velocities appear uncorrelated along the direction of propagation, which is consistent with the $\beta_G\sim1/2$ growth exponent (e.g.,\cite{fractal_surface}).

\subsection{Avalanches size and shape\label{sec:aval}}

\begin{figure}
\centering
\includegraphics[width=1\linewidth]{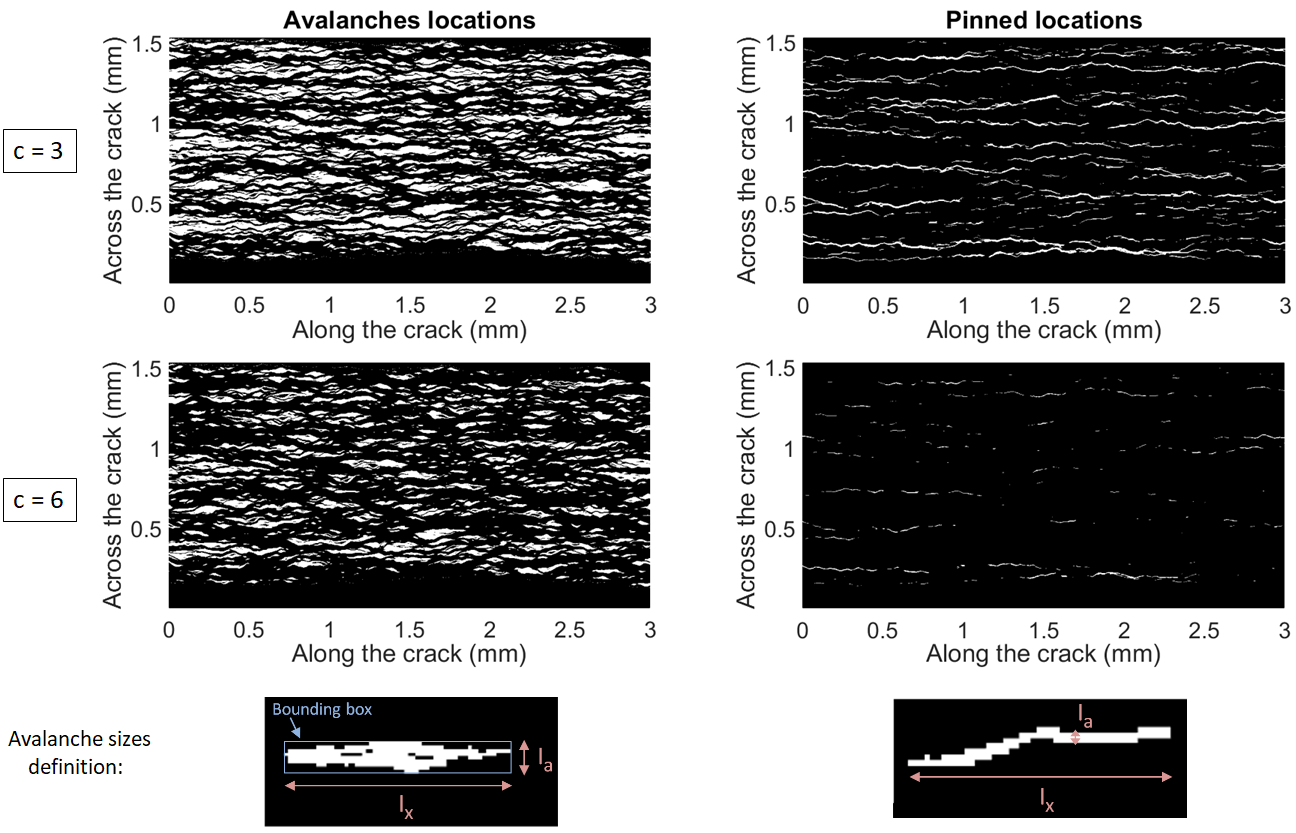}
  \caption{Positions of the avalanches (left) and pinning locations (right) in the front local velocity map $V(x,a)$ shown in Fig.\,\ref{fig:gc}c, as per Eqs.\,(\ref{threshold1}) and\,(\ref{threshold2}). Two thresholds are here used to define these maps relatively to the mean velocity: $c=3$ and $c=6$. The white areas are the locations of interest, of surfaces $S$, crossline extents $l_x$ and inline extents $l_a$. Bottom images: Difference in definition of $l_a$ for the avalanche (or depinning) and pinning clusters shown in Fig.\,\ref{fig:aval_map}. For the former, $l_a$ is the maximum extent along the $a$ direction. For the latter it is the average width in the same direction. In both cases, $l_x$ is the maximum extent along the $x$ direction and $S$ the full surface (in white) of the cluster. The square pattern marks the pixel size ($\Delta x=\Delta a = 3\,\upmu$m).}
 \label{fig:aval_map}
\end{figure}
\begin{figure}
\centering
\includegraphics[width=1\linewidth]{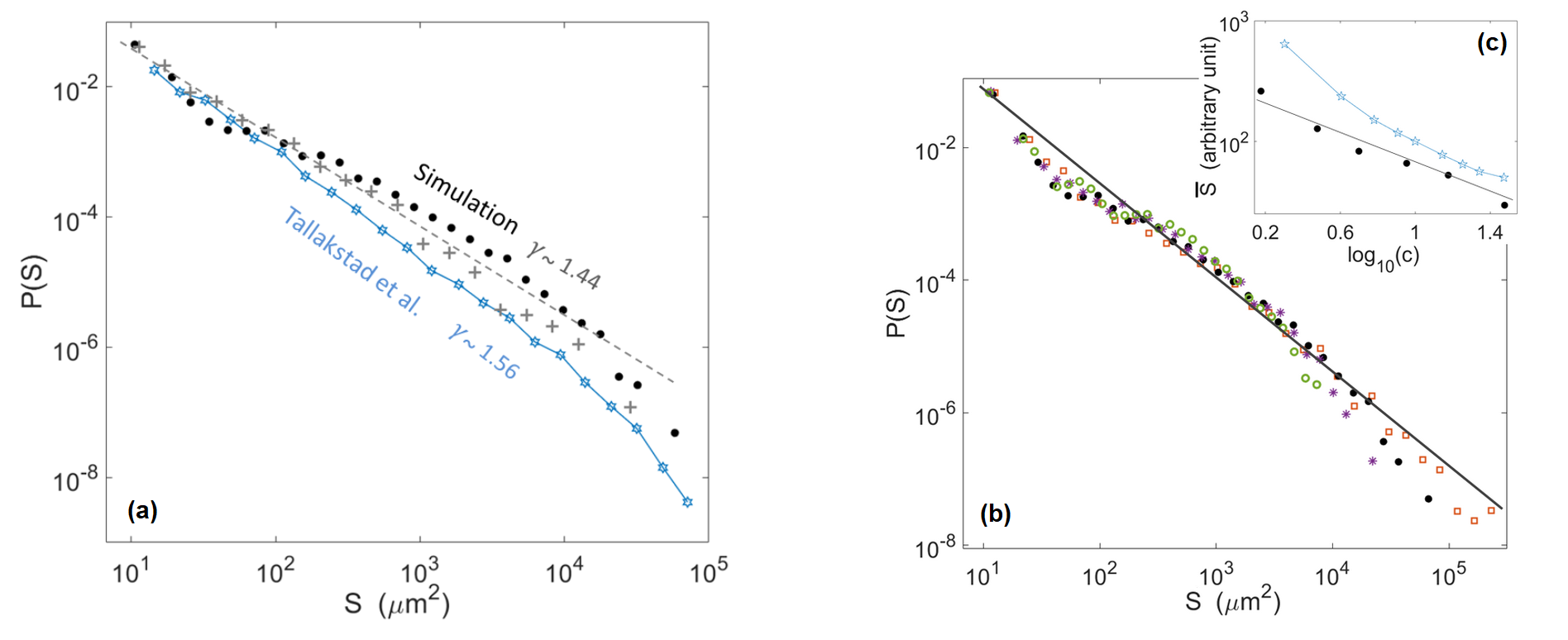}
  \caption{(a): Probability density function of the surface of the modelled avalanche clusters (plain points) and of the modelled pinning clusters (crosses), for a threshold $c=3$. The straight dashed line has a slope $\gamma=1.44$, as per Eq.\,(\ref{pseq}). For comparison, the hollow stars show the experimental probability density function obtained by Tallakstad \textit{et al.}\cite{Tallakstad2011} for the avalanche and pinning clusters (both are overlapping, see Fig.\,10 of their manuscript). (b): Same probability density function for various $c$ values: $c=1.5$ (squares), $c=3$ (plain points), $c=6$ (stars), $c=12$ (circles). The straight line has a slope $\gamma=1.4$, as per Eq.\,(\ref{pseq}). (c): Variation of the mean avalanche size $\overline{S}$ as a function of the threshold $c$ for the simulation (plain points) and the experiments (hollow stars). The modelled $\overline{S}$ is expressed in pixels (one pixel is $9\,\upmu$m\textsuperscript{2}) and the experimental $\overline{S}$ reported by Tallakstad \textit{et al.}\cite{Tallakstad2011} (on their Fig.\,13) is in an arbitrary unit, so that the magnitude of both should not here be compared. We have here shifted up this experimental data set for an easier comparison with the numerical one. The straight line has a slope $0.68$.}
 \label{fig:Psize}
\end{figure}
\begin{figure}
\centering
\includegraphics[width=1\linewidth]{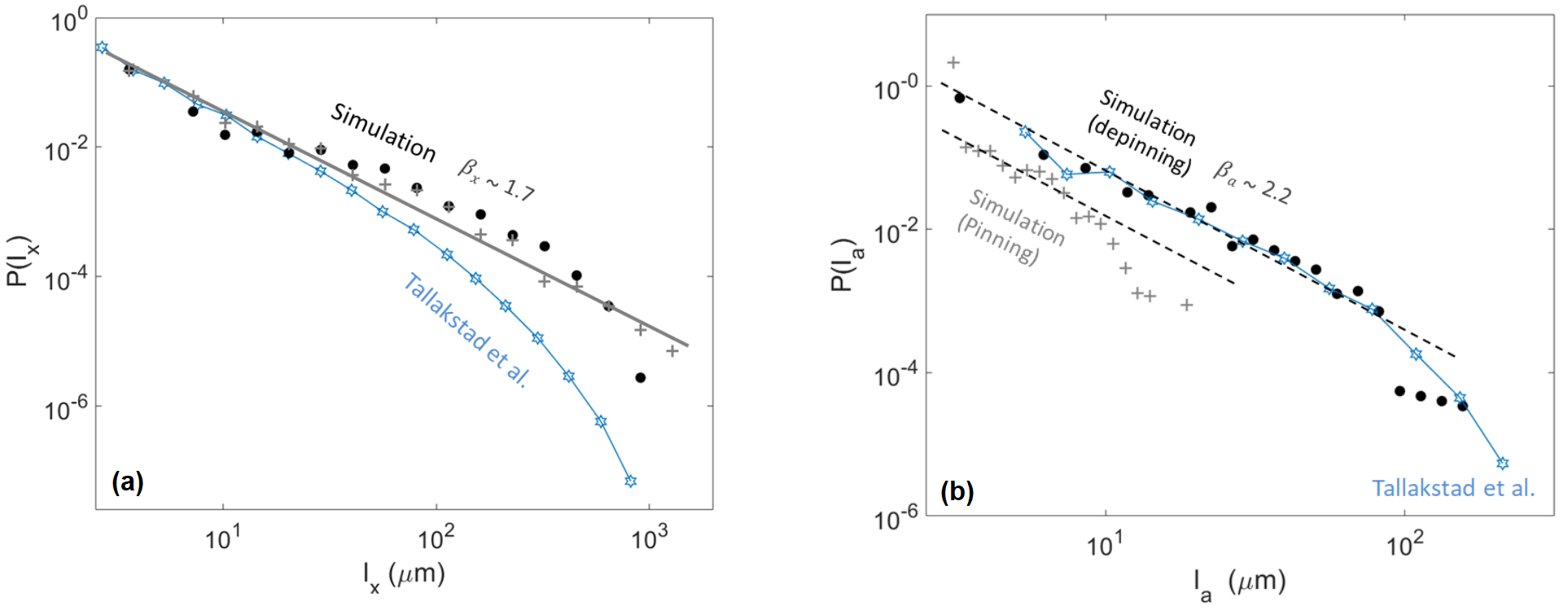}
  \caption{(a): Probability density function of the crossline extent $l_x$ of the modelled avalanche clusters (plain points) and of the modelled pinning clusters (crosses), for a threshold $c=3$. The straight line has a slope $\beta_x=1.7$, as per Eq.\,(\ref{plxeq}). The hollow stars shows the experimental probability density function obtained by Tallakstad \textit{et al.}\cite{Tallakstad2011} for the pinning and avalanche clusters (from their Fig.\,16a, inset, c=3). (b): Probability density function of the inline extent $l_a$ of the modelled avalanche clusters (plain points) and of the modelled pinning clusters (crosses), for a threshold $c=3$. The two straight dashed lines have a slope $\beta_x=2.2$, inline with that of the experimental data from Tallakstad \textit{et al.}\cite{Tallakstad2011} for the pinning clusters (hollow stars, from their Fig.\,16b, inset, c=3).}
 \label{fig:plx}
\end{figure}
We pursue by characterising the intermittent, burst-like, dynamics of our crack fronts and, more specifically, the avalanche (or depinning) and pinning clusters shown by the local front velocity $V(x,a)$. We define an avalanche when the front velocity locally exceeds the mean velocity $\overline{V}$ by an arbitrary threshold that we denote $c$, that is, when
\begin{equation}
   V(x,a)>c\overline{V}.
   \label{threshold1}
\end{equation}
Similarly, we state that a front is pinned when
\begin{equation}
   V(x,a)<\frac{\overline{V}}{c}.
   \label{threshold2}
\end{equation}
We then map, in Fig.\,\ref{fig:aval_map}, the thus defined avalanching and pinned locations of the crack. Following the analysis of Tallakstad \textit{et al.}\cite{Tallakstad2011}, we compute for each of these clusters the surface $S$, the crossline extent $l_x$ (that is, the maximum of a cluster width in the $x$ direction) and the inline extent $l_a$. The definition chosen for $l_a$ varies for the avalanche clusters, where the maximal extent along the $a$ direction is regarded, or the pinned one, where the mean extent along the $a$ direction is rather used. This choice was made\cite{Tallakstad2011} because the pinning clusters tend to be more tortuous so that their maximum span along the crack direction of propagation is not really representative of their actual extent (see Fig.\,\ref{fig:aval_map}).\\
In Fig.\,\ref{fig:Psize}a, we show the probability density function of the cluster surface $P(S)$ and compare it to the experimental one. One can notice that it behaves as
\begin{equation}
   P(S)\propto S^{-\gamma},
   \label{pseq}
\end{equation}
with $\gamma=1.44\pm0.15$. This value is comparable to the exponent inverted experimentally\cite{Tallakstad2011}, that is, $\gamma=1.56\pm0.04$.\\
Of course, the size of the avalanche (depinning) clusters highly depends on the chosen threshold $c$, but we verified, as experimentally reported, that the value of $\gamma$ inverted from the simulated data is not dependent on $c$, as shown in Fig.\,\ref{fig:Psize}b. We also show, in Fig\,\ref{fig:Psize}c, that the mean cluster size $\overline{S}$ varies with $c$ approximately as $\overline{S}\propto c^{-m}$, which $m\sim{0.68}$. This value is comparable with the experimental scaling law\cite{Tallakstad2011} measured to be $\overline{S}\propto c^{-0.75}$.
\\We also computed the probability density function of $l_x$ and $l_a$, that are respectively compared to their experimental equivalent in Fig.\,\ref{fig:plx}. These functions can be fitted with
\begin{equation}
   P(l_x)\propto {l_x}^{-\beta_x},
   \label{plxeq}
\end{equation}
\begin{equation}
   P(l_a)\propto {l_a}^{-\beta_a},
   \label{plaeq}
\end{equation}
and we found $\beta_x=1.7\pm0.2$, close to the reported experimental value\cite{Tallakstad2011} $\beta_x\sim1.93$. The value we found for $\beta_a\sim2.2=0.2$ is also inline with that of Tallakstad \textit{et al.}\cite{Tallakstad2011}, who reported $\beta_a\sim2.36$.\\
It should be noted that, while we have here fitted $P(S)$, $P(l_x)$ and $P(l_a)$ with plain scaling laws (i.e., with Eqs.\,(\ref{pseq}) to\,(\ref{plaeq})), Tallakstad \textit{et al.}\cite{Tallakstad2011} also studied the cut-off scales above which these scaling laws vanish in the experimental data, and the dependence of these cut-off scales with the arbitrary threshold $c$. In our case, such scales are challenging to define, as one can for instance notice in Figs.\,\ref{fig:Psize} and\,\ref{fig:plx}, where an exponential cut-off is not obvious. This may result from a limited statistical description of the larger avalanches in our simulations. Similar cut-off scales, decreasing with increasing $c$ should however hold in our numerical data, in order to explain the decrease of average avalanche size with $c$, as shown in Fig.\,\ref{fig:Psize}c.

\FloatBarrier
\subsection{Front morphology\label{sec:hurst}}

\begin{figure}
\centering
\includegraphics[width=1\linewidth]{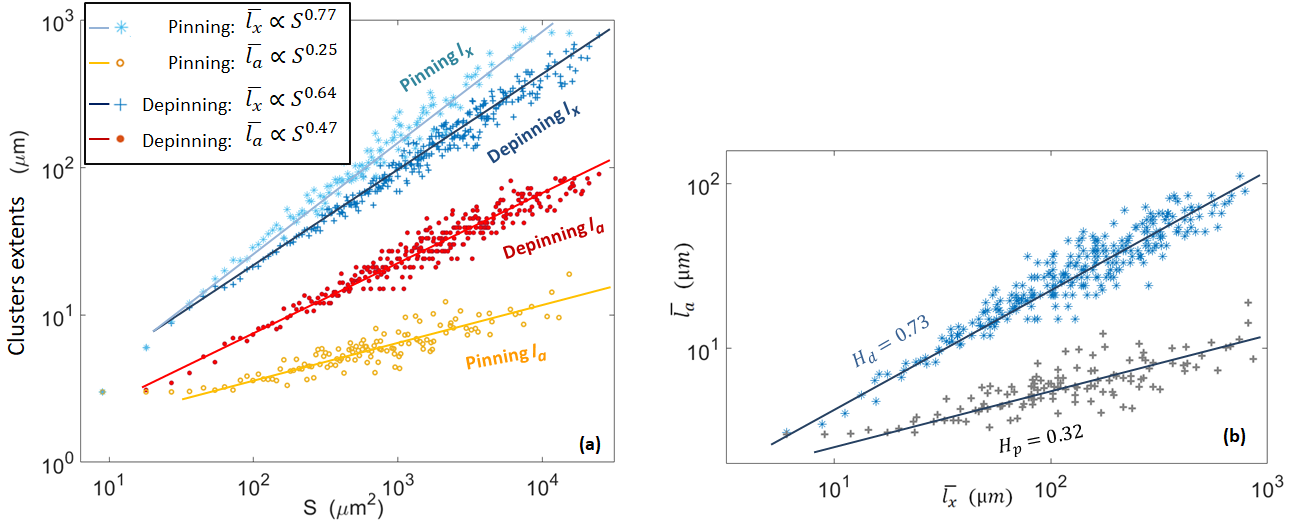}
  \caption{(a): Mean linear extents of the simulated pinning and depinning clusters as a function of cluster size. The four data sets are, from top to bottom, $\overline{l_x}$ for the pinning clusters (hollow stars), $\overline{l_x}$ for the avalanche clusters (crosses), $\overline{l_a}$ for the avalanche clusters (plain points), $\overline{l_a}$ for the pinning clusters (hollow points). The straight lines correspond to the fits described in the inset. See text for the equivalent experimental exponents. (b): Mean inline extent $\overline{l_a}$ as a function of the mean crossline extent $\overline{l_x}$ for the pinning and depinning clusters. The straight lines have a slope of, respectively, $H_p=0.32$ and $H_d=0.73$.}
 \label{fig:svsl}
\end{figure}

Finally, we show, in Fig.\,\ref{fig:svsl}\,a, the relations between the clusters surface $S$ and their linear extent $\overline{l_x}$ and $\overline{l_a}$. Here, $\overline{l_x}$ and $\overline{l_a}$ are the mean extents for all the observed clusters sharing a same surface (with the given pixel size). We could fit these relations with $\overline{l_x}\propto S^{0.77}$ and $\overline{l_a}\propto S^{0.25}$ for the pinning clusters, and with $\overline{l_x}\propto S^{0.64}$ and $\overline{l_a} \propto S^{0.47}$ for the avalanches clusters. It is in qualitative agreement with the laws observed by Tallakstad \textit{et al.}\cite{Tallakstad2011}: $\overline{l_x}\propto S^{0.63}$ and $\overline{l_a}\propto S^{0.34}$ for the pinning clusters, and $ S\propto \overline{l_x}^{0.61}$ and $\overline{l_a}\propto S^{0.41}$ for the avalanches clusters. These exponents were experimentally reported with a $\pm0.05$ accuracy, and we estimated comparable error bars for the numerically derived ones. Thus, the shape of our simulated avalanches and pinned locations is rather similar to the observed experimental ones.
Note that, from all the previous exponents, one can easily define $H$ such that $l_a\propto{l_x}^H$, and we thus have $H_p\sim0.25/0.77=0.32\pm0.1$ and $H_d\sim0.47/0.64=0.73\pm0.01$ for, respectively, the simulated pinning and depinning clusters (see Fig.\,\ref{fig:svsl}\,b).\\
It was suggested\cite{Laurson2010, confpaper} that $H$ is a good indicator of the front morphology, as the front shape is to be highly dependent on the aspect ratio of its avalanches. To verify this hypothesis, we computed the advancement fluctuation along the front $\sigma$, that is
\begin{equation}
   \sigma(\delta x)=\sqrt{{<}\left(a(x_0+\delta x, t)-a(x_0, t)\right)^2{>}_{x_0,t}}.
   \label{hurst}
\end{equation}
\begin{figure}
\centering
\includegraphics[width=1\linewidth]{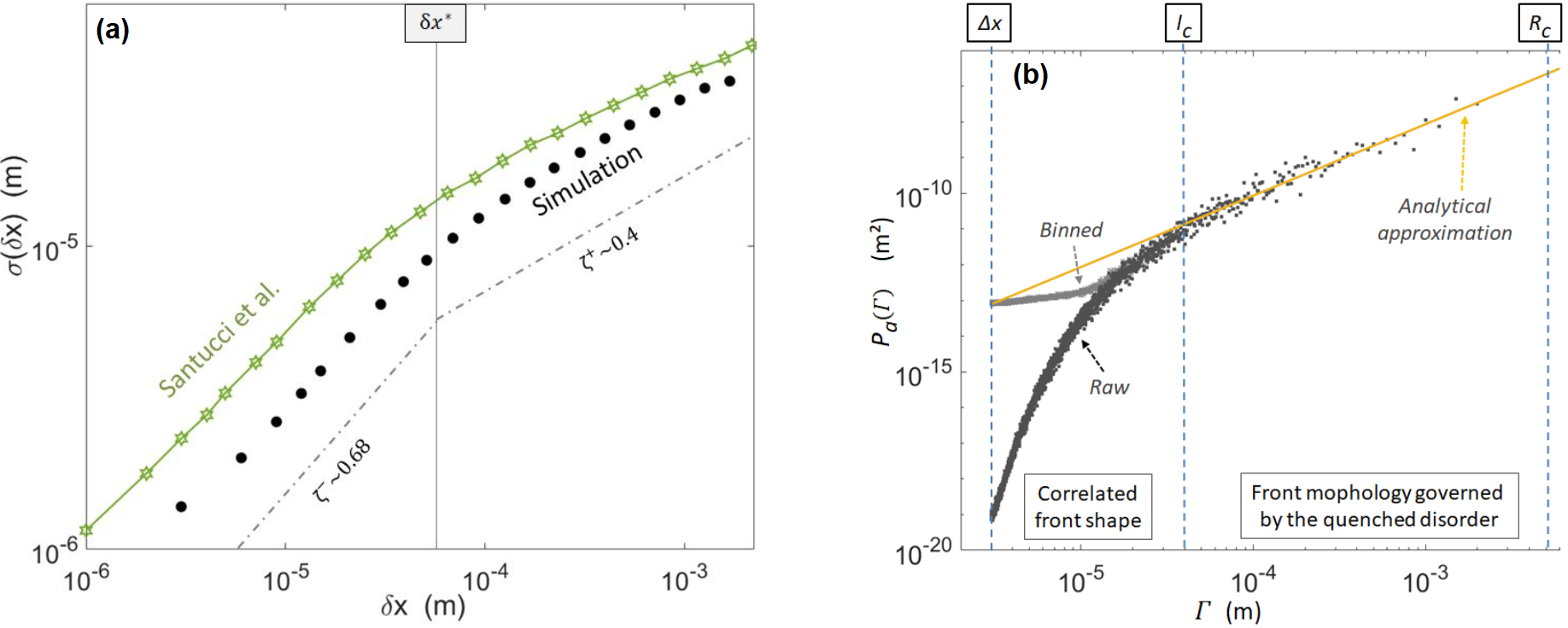}
  \caption{(a): Advancement fluctuation $\sigma$ along the crack fronts, as per Eq.\,(\ref{hurst}), for the simulation (plain points) and an experimental data set from Santucci \textit{et al.}\cite{Santucci_2010} (see their Fig.\,4). Different self-affine behaviours are observed above and below the $\delta x^*$ cut-off, with comparable Hurst exponents $\zeta$. The dashed lines mark the slopes fitted on the simulation data for the two cases. The experimental points are from an experiment different from those of Tallakstad \textit{et al.}\cite{Tallakstad2011} to which the model was calibrated. (b): Power spectra of the simulated crack advancement, averaged over $10,000$ consecutive fronts. It is shown both before (raw) and after (binned) binning the fronts to the experimental camera pixel size. The difference between these two plots shows an influence of the observation scale on the small-scale study of the crack morphology. The plain line is the approximation\cite{Cochard} from Eq.\,(\ref{powerspectra}), which is valid between $l_c$ and $R_c$, where the morphology is dominated by the material quenched disorder. Note that the scaling regime for scales above $l_c$ was already studied by Cochard \textit{et al.}\cite{Cochard}, while the model match to the experiment below this cut-off scale, shown in (a), is a new result, as already discussed in section\,\ref{sec:vel}. 
  }
 \label{fig:hurst}
\end{figure}
While this quantity was not presented by Tallakstad \textit{et al.}\cite{Tallakstad2011}, it was provided by other experimental works done on the same set-up\cite{Maloy2006,Santucci_2010}, and Fig.\,\ref{fig:hurst}a shows $\sigma$ as reported by these authors, together with that computed in the output of our simulation. One can notice that the numerical fronts are less rugous than the experimental ones.
Such a mismatch is here due to the fact that the experiment from Santucci \textit{et al.}, shown in Fig.\,\ref{fig:hurst}a, had more rugous crack fronts than the one from Tallakstad \textit{et al.}, to which the simulation is calibrated (as shown in Fig.\,\ref{fig:growth}). In both cases, the data sets seem to present two self-affine behaviours (e.g.,\cite{fractal_surface}) with a Hurst exponent $\zeta$ that differs at low and high length scales. Noting $\delta x^*$ the cut-off between these length scales we indeed have:
\begin{equation}
   \sigma\propto {\delta_x}^{\zeta^-}\,\,\,\text{for}\,\,\,\delta x < \delta x^*,
   \label{etam}
\end{equation}
\begin{equation}
   \sigma\propto {\delta_x}^{\zeta^+}\,\,\,\text{for}\,\,\,\delta x > \delta x^*.
   \label{etap}
\end{equation}
We derived $\zeta^-\sim0.68\pm0.05$ and $\zeta^+=0.4\pm0.05$ for the simulation, which compare well to the exponents that were measured experimentally, respectively, $\zeta^-=0.60\pm0.05$ and $\zeta^+=0.35\pm0.05$ and which are also close to the values we found for $H_d$ and $H_p$. The cut-off scale between the two regimes is also similar in both the experimental and numerical cases: $\delta x^*\sim80\,\upmu$m, comparable to the disorder correlation length $l_c$, and to the length scales $x^*$, below which the local propagation velocities are correlated.\\
For scales above this correlation length, Cochard \textit{et al.}\cite{Cochard} showed, by analytically analysing the same model as we here study, that the front morphology is dominated by the material quenched disorder with a Hurst coefficient approximating to $\zeta^+=0.5$. At even larger scales, above $R_c\sim\pi l_c \alpha^2 \overline{G_c}/(k_B T_0)$, they also showed\cite{Cochard} that the roughness of the simulated cracks ceases to be governed by the quenched disorder but is rather dominated by the thermal (annealed) noise, with $\sigma$ decaying logarithmically and with a Hurst coefficient tending to $\zeta^\infty=0$. With our set of parameters, $R_c$ computes to $6\,$mm, which is close to, yet bigger than, the total analysed length of the front. The value $\zeta^+\sim0.4$, that we have here inverted, arises then likely from the transition between the two regimes, $\zeta^+=0.5$ and $\zeta^\infty=0$, as already mentioned for the experimental case, in Ref.\cite{Santucci_2010}. In addition to a theoretical Hurst exponent $\zeta^+=0.5$, Cochard \textit{et al.}\cite{Cochard} computed an analytical approximation for the fronts morphology power spectrum $P_a(\Gamma)$, for the length scales $\Gamma$ for which the effect of the quenched disorder prevails:
\begin{equation}
P_a(\Gamma)\sim\left(\frac{\delta G_c}{\overline{G}}\right)^2\frac{\Gamma R_c}{4\pi^2}.
\label{powerspectra}
\end{equation}
We show, in Fig.\,\ref{fig:hurst}b, how this approximation also fits the power spectra of our modelled front.

\section{Discussion and conclusion}

\begin{table}
\centering
\begin{tabular}{c|c c c c c}
\textbf{\,Parameter\,}  & \textbf{\,\,\,\,\,\,Expt.\,\,\,\,\,\,}  & &\textbf{\,Models\,}& \\
 &  & \textit{\,\,\,\,\,ABL\,\,\,\,\,} & \textit{\,\,\,FB\,\,\,} & \textit{\,\,NSL\,\,}\\ \hline
$\beta_G$               & $0.55$ & $0.6$ & $0.52$ & \\ 
$\eta$       & $2.6$ & $2.6$ & $2.56$ & \\ 
$\tau_x$    & $0.53$    & $0.13$ & $0.4$ &         \\ 
$x^*$    & $\sim100\,\upmu$m    & $94\,\upmu$m &  &         \\ 
$\beta_x$    & $1.94$    & $1.7$ &  &         \\ 
$\beta_a$    & $2.34$    & $2.2$ &  &         \\ 
$\gamma$    & $1.56$    & $1.4$ &  & $1.5$        \\
$m$    & $0.75$    & $0.68$ &   &         \\
$\zeta^-$    & $0.60$    & $0.68$ & $0.67$ &  $0.48$       \\ 
$\zeta^+$    & $0.35$    & $0.4$ & $0.39$ & $0.37$  \\ 
$H_d$         &    $0.66$    &  $0.73$    &  $0.6$  &  $0.65$  \\
$H_p$         &  $0.55$    &   $0.32$    &   $0.4$   &    \\
\end{tabular}
\caption{Comparison of various exponents and cut-off scales derived experimentally\cite{Tallakstad2011, Santucci_2010} (Expt.) and numerically with the present, Arrhenius Based, fluctuating Line model (ABL), the Fiber Bundle model\cite{a_numerical_study} (FB) and the Non-Subcritical fluctuating Line model\cite{Laurson2010,Steph_PRL_crackling} (NSL).}
\label{tab_expos}
\end{table}
We studied an interfacial fracture propagation model, based only on statistical and subcritical physics in the sense of an Arrhenius law (Eq.\,(\ref{model1})) and on the elastic redistribution of stress along crack fronts (Eq.\,(\ref{model2})). Following the work of Cochard \textit{et al.}\cite{Cochard}, we here showed that it allows a good representation of the intermittent dynamics of fracture in disordered media, as it approximately mimics the scaling laws dictating the propagation of experimental fronts, such as their growth exponent, their local velocity distribution and space and time correlations, the size of their avalanches and their self-affine characteristics.\\
To run our simulations, we had to assume a given distribution for the toughness of the rupturing interface, as this quantity is not directly measurable in the laboratory. We proposed $G_c$ to be normally distributed with a unique correlation length and, of course, this can only be a rough approximation of the actual fracture energy obtained by Tallakstad \textit{et al.}\cite{Tallakstad2011} by sintering two sand-blasted plexiglass plates. From this approximation, could arise discrepancies between our simulations and the experiments. We have indeed shown how some of the observed exponents were strongly dependent on the definition of the material disorder. We also have assumed a perfectly elastic crack front, when the local dynamics of creeping PMMA could be visco-elastic in part, particularly below the typical length scale $r\,\sim\,GE/\sigma_y^2 \sim 30$\,$\upmu$m for plasticity around crack tips (e.g.,\cite{lawn_1993}) in this material, where $\sigma_y \sim 100$\,MPa is the tensile yield stress of the polymer and $E \sim 3$\,GPa its Young modulus\cite{altuglas}.\\
These points being stated, the vast majority of the statistical quantities that we have here studied show a good match to those from the experimental observations, so that both the considered physical model and the interface definition are likely to be relevant. A further validation of this thermally activated model could derive from the comparison of its predictions with interfacial experiments at various background temperatures $T_0$. However, such experimental data is, to our knowledge, not yet available. Of course, some of our considered parameters (e.g., $G_c$, $V_0$ or $\alpha$) may, in practice, be temperature dependent so that a straight transposition of the model to different background temperatures could prove to be too simple. Creep experiments in bulk PMMA at various room temperatures can however be found in the literature\cite{Marshall_1974}, where only the mean front velocity versus the mean mechanical load are measured. In this case\cite{Marshall_1974}, it is reported that the creep dynamics is compatible with an Arrhenius-like process. By submitting many different materials to a constant load, at various temperatures, their lifetime was also shown\cite{Zhurkov1984,Vanel_2009} to follow an Arrhenius law, with an energy barrier that decreases with the applied stress. These materials include metals, alloys, non-metallic crystals and polymers (and PMMA in particular).
\\It should be noted that, as stated in our introduction, other models have been considered to numerically reproduce the interfacial PMMA experiments, notably, a non-subcritical threshold based fluctuating line model by Tanguy \textit{et al.}\cite{Tanguy1998}, Bonamy \textit{et al.}\cite{Steph_PRL_crackling} or Laurson \textit{et al.}\cite{Laurson2010,avalanche_classes} and a fiber bundle approach by Schmittbuhl \textit{et al.}\cite{FBM_Schmittbuhl}, Gjerden \textit{et al.}\cite{a_numerical_study} or Stormo \textit{et al.}\cite{bundle2}. The present manuscript does not challenge these other models per se, but rather offers an alternative explanation to the intermittent propagation of rough cracks. The former model, the fluctuating line model\cite{Tanguy1998,Steph_PRL_crackling,Laurson2010,avalanche_classes}, considers a similar redistribution of energy release rate $G$ as proposed in Eq.\,(\ref{model2}), but with a dynamics that is thresholded rather than following a subcritical growth law. The fronts either move forward by one pixel\cite{Laurson2010} if $G>G_c$, or with a velocity proportional\cite{Steph_PRL_crackling} to ($G-G_c$). It is completely pinned otherwise ($V=0$ for $G<G_c$). While reproducing several statistical features of the experiments, this non-subcritical line propagation model does not simulate the mean propagation of cracks in various loading regimes (as done by Cochard \textit{et al.}\cite{Cochard}) or the distribution in local velocity\cite{Santucci_2018}, and, in particular, the power law tail of this distribution (i.e., Fig.\,\ref{fig:veldist}).
By contrast, the latter model\cite{FBM_Schmittbuhl,a_numerical_study,bundle2}, the fiber bundle one, can reproduce this particular power law tail. It is not a line model: the interface is sampled with parallel elastic fibers breaking at a given force threshold. This threshold is less in the vicinity of the crack than away from it (it is modelled with a linear gradient), explaining why the rupture is concentrated around a defined front, and it holds a random component in order to model the quenched disorder of the interface. An advantage of the fiber bundle model is to be able to describe a coalescence of damage in front of the crack\cite{bouchaud_coales} rather than solely describing a unique front. This could likely also be achieved in a subcritical framework, but would require to authorise damage in a full 2D plane, or require a full 3D modelling (i.e., also authorise out-of-plane damage), rather than only the modelling of a 2D front. In practice, thermal activation and damage coalescence may occur simultaneously. The observation of actual damage nucleation, in the experiments that we reproduce, has however never been obvious. Instead, the experimental fronts look rather continuous. Coalescence could yet still be at play at length scales smaller than the observation resolution. This being stated, an advantage of our model is to only rely on the experimental observations, on stress redistribution and on statistical physics.
Another clear advantage of the Arrhenius based model, when compared to the other ones, is to hold a subcritical description that is physically well understood and that is a good descriptor of creep in many materials\cite{lawn_1993,Vanel_2009}. For the record, we show in table\,\ref{tab_expos} a comparison between the different exponents predicted by the three models, that all successfully reproduce some experimental observables. \\
Note that, if linearizing Eq.\,(\ref{model1}) with a Taylor expansion around $\overline{G_c}-\overline{G}$, that is, for propagation velocities close to the mean crack speed $\overline{V}=V_0\exp(-\alpha^2[\overline{G_c}-\overline{G}]/[k_B T_0])$, one obtains
\begin{equation}
   V \sim V_\text{cst}-\frac{\alpha^2\overline{V}}{k_B T_0}(G_c-G),
   \label{linear}
\end{equation}
where $V_\text{cst}$ is a constant equal to $\overline{V}(1+\alpha^2[\overline{G_c}-\overline{G}]/[k_B T_0])$. This simplified form for our subcritical model is mathematically similar to that of the overcritical model (in the sense that a non zero velocity is only obtained for $G>G_c$) of Bonamy \textit{et al.}\cite{Steph_PRL_crackling}, where $V=\text{max}[\mu(G_c-G),0]$ and where the coefficient of proportionality $\mu$ was named the `effective mobility of the crack front'. Equation\,(\ref{linear}) may give some insight in the physical meaning of $\mu$ in this alternative model\cite{Steph_PRL_crackling}. While the above similitude in mathematical forms may explain the obtention of some similar exponents in the dynamics of the two models (see table\,\ref{tab_expos}), Eq.\,(\ref{linear}) is only a crude approximation of our highly non-linear Arrhenius formalism, which, as discussed below, allows a more exhaustive description of the experimental intermittent creep dynamics. In our simulations, the exponential Arrhenius probability term, describing the crack velocity, ranges over more than three orders of magnitude while ($G_c-G$) ranges over less than two decades.\\
Continuing with the comparison of our model with pre-existing ones, we had, in our case, to calibrate the disorder to the experimental data, in particular to accurately reproduce the $\eta$ exponent, that is, to reproduce the fat tail of the crack velocity distribution. Paradoxically, this exponent, which is not accounted for by the other line model, has been found to be rather constant across different experiments and experimental set-ups. It could indicate that, in practice, the disorder obtained experimentally from the blasting and sintering of PMMA plates has always been relatively similar. Such qualitative statement is of course difficult to verify, because there exists no direct way of measuring the fracture energy of the experimental sintered samples. From Fig.\,\ref{fig:veldist}b, one can yet notice that the calibration of the disorder amplitude does not need to be particularly accurate to obtain a qualitative fit to the experimental velocity distribution. The spread of the $\eta$ exponent, for large disorders, is not that important in our model for the range of considered $\delta G_c$, which can also be seen in Fig.\,5 of Cochard \textit{et al.}\cite{Cochard}. Gjerden \textit{et al.}\cite{fiber_universel} suggested that the nucleation of damages, predicted by their fiber bundle model, led to a new - percolation - universality class for the propagation of cracks, explaining in particular the robustness of the exponent $\eta$. Their studies are however also numerical and cover a finite range of disorders, and an extra analytical proof would be needed to show that a system of infinite size would lead exactly to the same exponent, for any disorder distribution shape and amplitude.\\
Despite the variety in models reproducing the rough dynamics of creep, the present work provides additional indications that a thermodynamics framework in the sense of a thermally activated subcritical crack growth is well suited for the description of creeping cracks. Such a framework has long been considered (e.g.,\cite{Brenner, Zhurkov1984, Scorretti_2001, roux, Santucci2004, Vanel_2009}), and, additionally to the scaling laws that we have here presented, the proposed model was proven to fit many other observable features of the physics of rupture\cite{Lengline2011,Cochard,TVD1,TVD2}. It accurately recreates the mean advancement of cracks under various loading conditions\cite{Lengline2011, Cochard}, including when a front creeps in a spontaneous (not forced) relaxation regime, which cannot be achieved with the other (non subcritical) models, predicting an immobile front. When coupled with heat dissipation at the fracture tip, our description also accounts for the brittleness of matter\cite{TVD2} and for its brittle-ductile transition\cite{TVD1}.\\
Indeed, for zero dimensional (scalar) crack fronts, it was shown\cite{TVD2} that the thermal fluctuation at the crack tip, expressed as a deviation of the temperature from $T_0$ in Eq.\,(\ref{model1}), can explain the transition between creep and abrupt rupture, that is, the transition to a propagation velocity close to a mechanical wave speed $V_0$, five orders of magnitude higher than the maximal creep velocity $V$ that was here modelled. It was also shown, similarly to many phase transition problems, that such a thermal transition could be favoured by material disorder\cite{TVD1}. Thus, a direct continuation of the present work could be to introduce such a heat dissipation for interfacial cracks in order to study how brittle avalanches nucleate at given positions (typically positions with weaker $G_c$) to then expand laterally to become bulk threatening events.

\section*{Acknowledgement}

\noindent The authors declare no competing interests in the publishing of this work. They acknowledge the support of the Universities of Strasbourg and Oslo, of the CNRS INSU ALEAS program and of the IRP France-Norway D-FFRACT. We thank the Research Council of Norway through its Centres of Excellence funding scheme, project number 262644. We are also grateful for the support of the Lavrentyev Institute of Hydrodynamics, through Grant No 14.W03.31.0002 of the Russian Government.

\newpage
\section*{Appendices}
\appendix

\vspace{1cm}
\section{On the negligible temperature elevation\label{tempelev}}

In this manuscript, we have considered that the temperature elevation at the crack front $\Delta T$, arising from the release of energy $G$, was negligible. In this section, we discuss this assumption further. To assess $\Delta T$, We use a quasi-static thermal model that we have introduced in several previous works\cite{ToussaintSoft,TVD1,TVD2}. There, cracks having a velocity less than $\lambda/(\pi  Cl)$ experience a rise in temperature
\begin{equation}
  \Delta T \sim \frac{\phi GV}{\lambda},
  \label{maxcreepT}
\end{equation}
where $\phi$ is a heat conversion efficiency as the front advances, where $\lambda=0.18$\,J\,s\textsuperscript{-1}\,m\textsuperscript{-1}\,K\textsuperscript{-1} and $C=1.5$\,MJ\,m\textsuperscript{-3} denote, respectively, the heat conductivity and capacity of PMMA\cite{altuglas}, and where $l$ is the radius of a heat production zone around the front. 
\\In the case of bulk PMMA, it was inverted\cite{TVD2} that $\phi\sim0.2$ and $l\sim10\,$nm. Of course, applying these bulk PMMA thermal parameters to the rupture dynamics of sintered interfacial PMMA is already an assumption, as both are, in practice, two different (although similar) materials. With this assumption, Eq.\,(\ref{maxcreepT}) is valid for $V < \lambda/(\pi  Cl) \sim 4\,$m\,s\textsuperscript{-1}, that is, it is valid for any velocity that we have modelled, which is less than $1\,$mm\,s\textsuperscript{-1} (see Fig.\,\ref{fig:gc}). With this conservative $V=1\,$mm\,s\textsuperscript{-1} and with a maximal load of about $250\,$J\,m\textsuperscript{-2} (see Fig.\,\ref{Gdistri}a), the maximal temperature elevation $\Delta T$ computes to less than a degree and, thus, is negligible compared to $T_0=298\,$K. In this context, our adiabatic crack front hypothesis is verified.

\vspace{1cm}
\section{On the physical meaning of the $\alpha$ parameter\label{alpha}}
The $\alpha^2$ parameter is a particularly small (subatomic) area. This parameter was directly fitted on the creep curve of the studied interfacial PMMA, showing an exponential dependence of the average front velocity with the mean crack energy release rate\cite{Lengline2011}. In Refs.\cite{Vanel_2009,TVD2,TVD3}, we explain how $\alpha$ is not an actual physical size of the rupturing material. Instead, $\alpha^2$ is an equivalent area in the order of $d_0^3/l$, where $d_0$ is the typical intra-molecular distance (called for in a thermally activated context) and where $l$ is the scale limiting the stress divergence – and thus the local storage of energy – at the crack tip. Assuming $d_0$ to be about $0.3$ nanometers, one gets $l$ in the order of 1-10 nanometers. This size is orders of magnitude smaller than the typical process zone length around crack tips in PMMA ($10$ to $100$ micrometers). There is, however, no strong reason to consider a Dugdale\cite{dugdale1960} description of the stress in the process zone, that is, that the stress saturates at scales below $10$ to $100$ micrometers from the front. It is instead likely that the stress remains an increasing function up to small scales inside the process zone. In PMMA, a few nanometers (i.e., $l$) is a physically reasonable length scale, as it is both the size of a few MMA radicals and a typical length scale for the entanglement density\cite{Entangled} in the polymer.

\vspace{1cm}
\section{Solver convergence\label{conv}}

\begin{figure}
\centering
\includegraphics[width=1\linewidth]{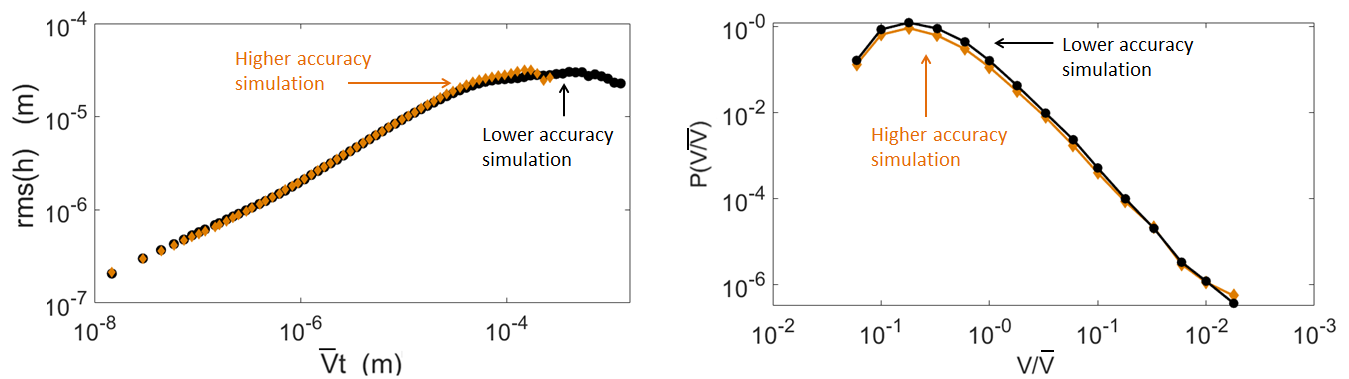}
  \caption{Front width growth (a) and local velocity distribution (b) for two simulations with the same physical parameters but different numerical accuracies, as per table\,\ref{table_acc}. The points are data from the simulation analysed in this manuscript and the squares were computed on denser numerical grids. The computed exponents are not significantly affected. The most refined (heavier to run) simulation was run on a shorter time span, which explains why $\overline{V}t$ stops at about $2 \times 10^4$\,m.}
 \label{fig:acc}
\end{figure}
We verified that our simulations were accurate enough so that the derived statistical features were not dependent on the steps of the numerical computation grids. In table\,\ref{table_acc}, we show the accuracy parameters of two different simulations and show, in Fig.\,\ref{fig:acc}, how the modelled crack dynamics is unchanged with both parameter sets.
\begin{table}[ht]
\centering
\begin{tabular}{c|c c c}
\textbf{\,Parameter\,} & \textbf{\,Higher accuracy\,} & \textbf{\,Lower accuracy\,} &  \textbf{\,Unit\,}\\ \hline
$\Delta x_s$     & $0.5$  &  $1$ &  $\upmu$m     \\
$\Delta t_s$     & $\sim1$ & $\sim5$ &  ms     \\
$L_s$     & $12000$  & $6000$ &   $\upmu$m     \\
\end{tabular}
\caption{Two different sets of numerical accuracy parameters corresponding to the results shown in Fig.\,\ref{fig:acc}.}
\label{table_acc}
\end{table}

\vspace{1cm}
\section{On the time dependency of $\overline{G}$\label{gconst}}

\begin{figure}
\centering
\includegraphics[width=1\linewidth]{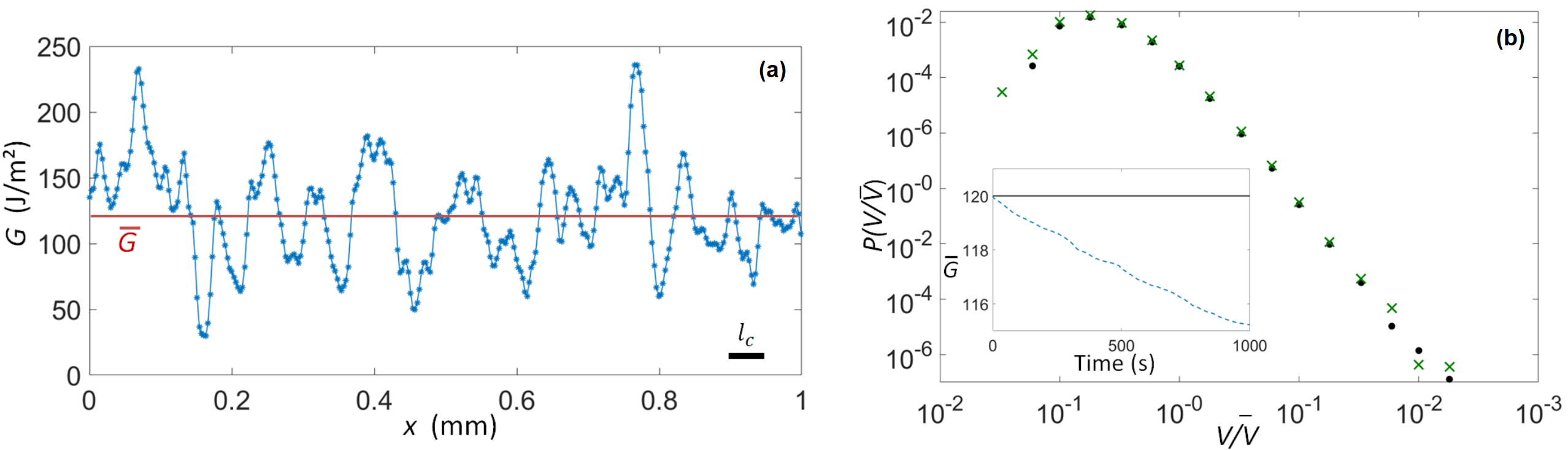}
  \caption{(a): Lateral variation of the energy release rate $G$ along a portion of the crack front, at a given simulation time, as predicted by Eq.\,(\ref{model2}) and with the parameters of table\,\ref{tab_param}. The straight line shows $\overline{G}=120\,$J\,m\textsuperscript{-2}. The standard deviation of $G$ computes to about $30$\% of this average value. For reference, the scale bar shows the correlation length $l_c=50\,\upmu$m of the material disorder. (b): Local velocity distribution for a simulation with a constant load $\overline{G}$, as in the manuscript (points), and with a relaxing load as the crack advances (crosses), as per Eq.\,(\ref{GmodeI}). The inset shows the time evolution of $\overline{G}$ for the former (plain line) and the latter (dashed line). The load relaxation during the second simulation stays small if compared to its lateral variation, as shown in (a).}
 \label{Gdistri}
\end{figure}

We have considered $\overline{G}$, in Eq.\,(\ref{model2}), to be a constant when running our simulations. In practice, the mean energy release rate is to vary during the progression of a crack. In the set-up of Tallakstad \textit{et al.}\cite{Tallakstad2011} (shown in Fig.\,\ref{pmma_sep}), $\overline{G}$ increases with the lower plate's deflection, while it decreases with the mean advance of the front $\overline{a}$. Using the Euler–Bernoulli beam theory \cite{fracmec}, one can actually compute\cite{Lengline2011} the mean energy release rate at the tip of such a system:
\begin{equation}
\overline{G}(t)=\frac{3 Eh_p^3u(t)^2}{8{\overline{a}(t)}^4} \hspace{0.5cm} \text{if } \overline{a} \gg h_p,
\label{GmodeI}
\end{equation}
with $E$ the lower plate Young modulus, $h_p$ its thickness, and $u$ the plate deflection (in meter). Two loading conditions were used in the experiments that we have here reproduced. One corresponds to a forced regime, where $u$ increases linearly with time, and where it was shown\cite{Lengline2011} that the resulting $\overline{G}$ is rather constant. The other regime is a relaxation one, where $u$ is kept constant while the crack continues to creep. In both cases, the long term evolution of $\overline{G}$ was shown\cite{Lengline2011,Cochard} to be reproduced by Eqs.\,(\ref{model1}) and\,(\ref{GmodeI}). In the second (relaxation) regime, $\overline{G}$ decreases with time, by a percentage given by Eq.\,(\ref{GmodeI}): [$\overline{a}_0/(\overline{a}_0+\Delta \overline{a})]^4$, where $\overline{a}_0$ is the crack length at the beginning of an experiment (typically $10$\,cm) and $\Delta \overline{a}$ is the total crack advancement during a realisation, which is in the order of $1\,$mm, similarly to what is shown in Fig.\,\ref{fig:gc}. The mean energy release rate thus decreases by about $4$\% during a non-forced experiment, and even less so ($0.4$\%) during typical avalanches of extent less than $0.1$\,mm (see Fig.\,\ref{fig:plx}b). In comparison, the spatial standard deviation of the energy release, as predicted by Eq.\,(\ref{model2}) and as shown for a given time $t$ in Fig.\,\ref{Gdistri}a, accounts for about $30$\% of $\overline{G}$. The time evolution of $G$ is hence small compared to its spatial variations, and modelling $\overline{G}$ as a constant is thus appropriate to study the burst-like dynamics of the experimental\cite{Tallakstad2011} cracks. Note however that it is possible to include Eq.\,(\ref{GmodeI}), or any other mechanical load descriptor, in the numerical solver, as done with our model by Cochard \textit{et al.}\cite{Cochard}. In Fig.\,\ref{Gdistri}b, we show that the velocity distribution does not change significantly when using Eq.\,(\ref{GmodeI}) to describe a relaxing load as the crack advances over the typical experimental course, compared to the constant $\overline{G}$ case studied in this manuscript.

\vspace{1cm}
\section{Varying the modelled mean velocity\label{veldep}}

\begin{figure}[H]
\centering
\includegraphics[width=1\linewidth]{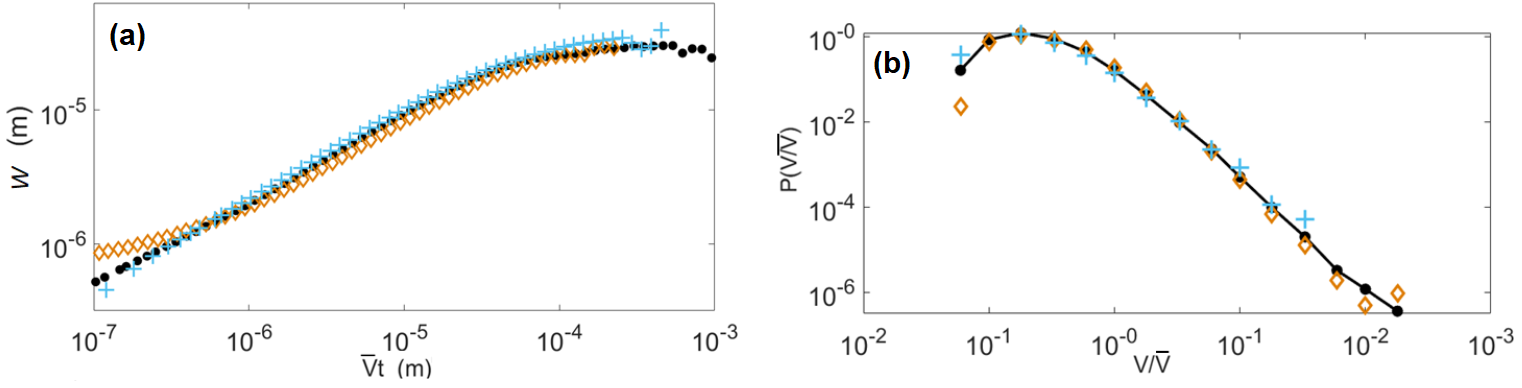}
  \caption{Front width growth (a) and local velocity distribution (b) for three simulations with a different average energy release rate: $\overline{G}=100$\,J\,m\textsuperscript{-2} (crosses), $\overline{G}=120$\,J\,m\textsuperscript{-2} (points) and $\overline{G}=140$\,J\,m\textsuperscript{-2} (squares). This corresponds to an average propagation velocity $\overline{V}$ of, respectively, $0.06\,\upmu$m\,s\textsuperscript{-1}, $1.5\,\upmu$m\,s\textsuperscript{-1} and $25\,\upmu$m\,s\textsuperscript{-1}. Both statistics show little dependency on $\overline{V}$, which is consistent with the experimental results\cite{Tallakstad2011}.}
 \label{fig:veldep}
\end{figure}

In Fig.\,\ref{fig:veldep}, we show that the intermittent dynamics of the simulated fronts is not strongly dependent on the average propagation velocity of the crack. This is consistent with the experimental observations from Tallakstad \textit{et al.}\cite{Tallakstad2011}, where many driving velocities were used.


\FloatBarrier

\begin{thebibliography}{10}
\urlstyle{rm}
\expandafter\ifx\csname url\endcsname\relax
  \def\url#1{\texttt{#1}}\fi
\expandafter\ifx\csname urlprefix\endcsname\relax\def\urlprefix{URL }\fi
\expandafter\ifx\csname doiprefix\endcsname\relax\def\doiprefix{DOI: }\fi
\providecommand{\bibinfo}[2]{#2}
\providecommand{\eprint}[2][]{\url{#2}}

\bibitem{lawn_1993}
\bibinfo{author}{Lawn, B.}
\newblock \emph{\bibinfo{title}{Fracture of Brittle Solids}}.
\newblock Cambridge Solid State Science Series (\bibinfo{publisher}{Cambridge
  University Press}, \bibinfo{year}{1993}), \bibinfo{edition}{2} edn.

\bibitem{porosity_crack_initiation}
\bibinfo{author}{Gerard, D.} \& \bibinfo{author}{Koss, D.}
\newblock \bibinfo{journal}{\bibinfo{title}{Porosity and crack initiation
  during low cycle fatigue}}.
\newblock {\emph{\JournalTitle{Materials Science and Engineering: A}}}
  \textbf{\bibinfo{volume}{129}}, \bibinfo{pages}{77 -- 85},
  \doiprefix\url{10.1016/0921-5093(90)90346-5} (\bibinfo{year}{1990}).

\bibitem{GaoRice_redistribution}
\bibinfo{author}{Gao, H.} \& \bibinfo{author}{Rice, J.~R.}
\newblock \bibinfo{journal}{\bibinfo{title}{{A First-Order Perturbation
  Analysis of Crack Trapping by Arrays of Obstacles}}}.
\newblock {\emph{\JournalTitle{Journal of Applied Mechanics}}}
  \textbf{\bibinfo{volume}{56}}, \bibinfo{pages}{828--836},
  \doiprefix\url{10.1115/1.3176178} (\bibinfo{year}{1989}).

\bibitem{Steph_PRL_crackling}
\bibinfo{author}{Bonamy, D.}, \bibinfo{author}{Santucci, S.} \&
  \bibinfo{author}{Ponson, L.}
\newblock \bibinfo{journal}{\bibinfo{title}{Crackling dynamics in material
  failure as the signature of a self-organized dynamic phase transition}}.
\newblock {\emph{\JournalTitle{Phys. Rev. Lett.}}}
  \textbf{\bibinfo{volume}{101}}, \bibinfo{pages}{045501},
  \doiprefix\url{10.1103/PhysRevLett.101.045501} (\bibinfo{year}{2008}).

\bibitem{Laurson2010}
\bibinfo{author}{Laurson, L.}, \bibinfo{author}{Santucci, S.} \&
  \bibinfo{author}{Zapperi, S.}
\newblock \bibinfo{journal}{\bibinfo{title}{Avalanches and clusters in planar
  crack front propagation}}.
\newblock {\emph{\JournalTitle{Phys. Rev. E}}} \textbf{\bibinfo{volume}{81}},
  \bibinfo{pages}{046116}, \doiprefix\url{10.1103/PhysRevE.81.046116}
  (\bibinfo{year}{2010}).

\bibitem{FBM_Schmittbuhl}
\bibinfo{author}{Schmittbuhl, J.}, \bibinfo{author}{Hansen, A.} \&
  \bibinfo{author}{Batrouni, G.~G.}
\newblock \bibinfo{journal}{\bibinfo{title}{Roughness of interfacial crack
  fronts: Stress-weighted percolation in the damage zone}}.
\newblock {\emph{\JournalTitle{Phys. Rev. Lett.}}}
  \textbf{\bibinfo{volume}{90}}, \bibinfo{pages}{045505},
  \doiprefix\url{10.1103/PhysRevLett.90.045505} (\bibinfo{year}{2003}).

\bibitem{Crackling_fiber_bundle}
\bibinfo{author}{Danku, Z.}, \bibinfo{author}{Kun, F.} \&
  \bibinfo{author}{Herrmann, H.~J.}
\newblock \bibinfo{journal}{\bibinfo{title}{Fractal frontiers of bursts and
  cracks in a fiber bundle model of creep rupture}}.
\newblock {\emph{\JournalTitle{Phys. Rev. E}}} \textbf{\bibinfo{volume}{92}},
  \bibinfo{pages}{062402}, \doiprefix\url{10.1103/PhysRevE.92.062402}
  (\bibinfo{year}{2015}).

\bibitem{Cochard}
\bibinfo{author}{Cochard, A.}, \bibinfo{author}{Lengliné, O.},
  \bibinfo{author}{Måløy, K.~J.} \& \bibinfo{author}{Toussaint, R.}
\newblock \bibinfo{journal}{\bibinfo{title}{Thermally activated crack fronts
  propagating in pinning disorder: simultaneous brittle/creep behavior
  depending on scale}}.
\newblock {\emph{\JournalTitle{Philosophical Transactions of the Royal Society
  A : Mathematical, Physical and Engineering Sciences}}}
  \doiprefix\url{10.1098/rsta.2017.0399} (\bibinfo{year}{2018}).

\bibitem{Wiese2021}
\bibinfo{author}{Wiese, K.~J.}
\newblock \bibinfo{title}{Theory and experiments for disordered elastic
  manifolds, depinning, avalanches, and sandpiles} (\bibinfo{year}{2021}).
\newblock \bibinfo{note}{Preprint, arXiv 2102.01215}.

\bibitem{Griffith1921}
\bibinfo{author}{Griffith, A.}
\newblock \bibinfo{journal}{\bibinfo{title}{{The Phenomena of Rupture and Flow
  in Solids}}}.
\newblock {\emph{\JournalTitle{Philosophical Transactions of the Royal Society
  of London A: Mathematical, Physical and Engineering Sciences}}}
  \textbf{\bibinfo{volume}{221}}, \bibinfo{pages}{163--198},
  \doiprefix\url{10.1098/rsta.1921.0006} (\bibinfo{year}{1921}).

\bibitem{Irwin1957}
\bibinfo{author}{Irwin, G.~R.}
\newblock \bibinfo{journal}{\bibinfo{title}{{Analysis of stresses and strains
  near the end of a crack traversing a plate}}}.
\newblock {\emph{\JournalTitle{Journal of Applied Mechanics}}}
  \textbf{\bibinfo{volume}{24}}, \bibinfo{pages}{361--364}
  (\bibinfo{year}{1957}).

\bibitem{Bares14}
\bibinfo{author}{Bar{\'e}s, J.}, \bibinfo{author}{Hattali, M.~L.},
  \bibinfo{author}{Dalmas, D.} \& \bibinfo{author}{Bonamy, D.}
\newblock \bibinfo{journal}{\bibinfo{title}{Fluctuations of global energy
  release and crackling in nominally brittle heterogeneous fracture}}.
\newblock {\emph{\JournalTitle{Physical Review Letters}}}
  \textbf{\bibinfo{volume}{113}},
  \doiprefix\url{10.1103/physrevlett.113.264301} (\bibinfo{year}{2014}).

\bibitem{crackling_crack2}
\bibinfo{author}{Turquet, A.~L.} \emph{et~al.}
\newblock \bibinfo{journal}{\bibinfo{title}{Source localization of microseismic
  emissions during pneumatic fracturing}}.
\newblock {\emph{\JournalTitle{Geophysical Research Letters}}}
  \textbf{\bibinfo{volume}{46}}, \bibinfo{pages}{3726--3733},
  \doiprefix\url{10.1029/2019GL082198} (\bibinfo{year}{2019}).

\bibitem{crackling_crack1}
\bibinfo{author}{Vu, C.-C.} \& \bibinfo{author}{Weiss, J.}
\newblock \bibinfo{journal}{\bibinfo{title}{Asymmetric damage avalanche shape
  in quasibrittle materials and subavalanche (aftershock) clusters}}.
\newblock {\emph{\JournalTitle{Phys. Rev. Lett.}}}
  \textbf{\bibinfo{volume}{125}}, \bibinfo{pages}{105502},
  \doiprefix\url{10.1103/PhysRevLett.125.105502} (\bibinfo{year}{2020}).

\bibitem{crackling_imbibition}
\bibinfo{author}{Santucci, S.}, \bibinfo{author}{Planet, R.},
  \bibinfo{author}{M{\aa}l{\o}y, K.~J.} \& \bibinfo{author}{Ort{\'{\i}}n, J.}
\newblock \bibinfo{journal}{\bibinfo{title}{Avalanches of imbibition fronts:
  Towards critical pinning}}.
\newblock {\emph{\JournalTitle{Europhysics Letters}}}
  \textbf{\bibinfo{volume}{94}}, \bibinfo{pages}{46005},
  \doiprefix\url{10.1209/0295-5075/94/46005} (\bibinfo{year}{2011}).

\bibitem{crackling_martensitic}
\bibinfo{author}{Vives, E.} \emph{et~al.}
\newblock \bibinfo{journal}{\bibinfo{title}{Distributions of avalanches in
  martensitic transformations}}.
\newblock {\emph{\JournalTitle{Phys. Rev. Lett.}}}
  \textbf{\bibinfo{volume}{72}}, \bibinfo{pages}{1694--1697},
  \doiprefix\url{10.1103/PhysRevLett.72.1694} (\bibinfo{year}{1994}).

\bibitem{crackling_plastic}
\bibinfo{author}{Dimiduk, D.~M.}, \bibinfo{author}{Woodward, C.},
  \bibinfo{author}{LeSar, R.} \& \bibinfo{author}{Uchic, M.~D.}
\newblock \bibinfo{journal}{\bibinfo{title}{Scale-free intermittent flow in
  crystal plasticity}}.
\newblock {\emph{\JournalTitle{Science}}} \textbf{\bibinfo{volume}{312}},
  \bibinfo{pages}{1188--1190}, \doiprefix\url{10.1126/science.1123889}
  (\bibinfo{year}{2006}).

\bibitem{crackling_Barkhausen}
\bibinfo{author}{Durin, G.} \& \bibinfo{author}{Zapperi, S.}
\newblock \bibinfo{journal}{\bibinfo{title}{Scaling exponents for {B}arkhausen
  avalanches in polycrystalline and amorphous ferromagnets}}.
\newblock {\emph{\JournalTitle{Phys. Rev. Lett.}}}
  \textbf{\bibinfo{volume}{84}}, \bibinfo{pages}{4705--4708},
  \doiprefix\url{10.1103/PhysRevLett.84.4705} (\bibinfo{year}{2000}).

\bibitem{crackling_nat}
\bibinfo{author}{Sethna, J.~P.}, \bibinfo{author}{Dahmen, K.~A.} \&
  \bibinfo{author}{Myers, C.~R.}
\newblock \bibinfo{journal}{\bibinfo{title}{Crackling noise}}.
\newblock {\emph{\JournalTitle{Nature}}} \doiprefix\url{10.1038/35065675}
  (\bibinfo{year}{2001}).

\bibitem{avalanche_classes}
\bibinfo{author}{Laurson, L.} \emph{et~al.}
\newblock \bibinfo{journal}{\bibinfo{title}{Evolution of the average avalanche
  shape with the universality class}}.
\newblock {\emph{\JournalTitle{Nature Communications}}}
  \bibinfo{pages}{242--250}, \doiprefix\url{10.1038/ncomms3927}
  (\bibinfo{year}{2013}).

\bibitem{crackling_fault_1}
\bibinfo{author}{Jolivet, R.} \emph{et~al.}
\newblock \bibinfo{journal}{\bibinfo{title}{{The Burst‐Like Behavior of
  Aseismic Slip on a Rough Fault: The Creeping Section of the Haiyuan Fault,
  China}}}.
\newblock {\emph{\JournalTitle{Bulletin of the Seismological Society of
  America}}} \textbf{\bibinfo{volume}{105}}, \bibinfo{pages}{480--488},
  \doiprefix\url{10.1785/0120140237} (\bibinfo{year}{2014}).

\bibitem{crackling_fault_2}
\bibinfo{author}{Rousset, B.} \emph{et~al.}
\newblock \bibinfo{journal}{\bibinfo{title}{An aseismic slip transient on the
  north anatolian fault}}.
\newblock {\emph{\JournalTitle{Geophysical Research Letters}}}
  \textbf{\bibinfo{volume}{43}}, \bibinfo{pages}{3254--3262},
  \doiprefix\url{10.1002/2016GL068250} (\bibinfo{year}{2016}).

\bibitem{crackling_fault_3}
\bibinfo{author}{Grob, M.} \emph{et~al.}
\newblock \bibinfo{journal}{\bibinfo{title}{Quake catalogs from an optical
  monitoring of an interfacial crack propagation}}.
\newblock {\emph{\JournalTitle{Pure and Applied Geophysics}}}
  \textbf{\bibinfo{volume}{166}}, \bibinfo{pages}{777--799},
  \doiprefix\url{10.1007/s00024-004-0496-z} (\bibinfo{year}{2009}).

\bibitem{crackling_fault_4}
\bibinfo{author}{Lengliné, O.} \emph{et~al.}
\newblock \bibinfo{journal}{\bibinfo{title}{Downscaling of fracture energy
  during brittle creep experiments}}.
\newblock {\emph{\JournalTitle{Journal of Geophysical Research: Solid Earth}}}
  \textbf{\bibinfo{volume}{116}}, \doiprefix\url{10.1029/2010JB008059}
  (\bibinfo{year}{2011}).

\bibitem{crackling_fault_5}
\bibinfo{author}{Lengliné, O.} \emph{et~al.}
\newblock \bibinfo{journal}{\bibinfo{title}{Interplay of seismic and aseismic
  deformations during earthquake swarms: An experimental approach}}.
\newblock {\emph{\JournalTitle{Earth and Planetary Science Letters}}}
  \textbf{\bibinfo{volume}{331-332}}, \bibinfo{pages}{215 -- 223},
  \doiprefix\url{10.1016/j.epsl.2012.03.022} (\bibinfo{year}{2012}).

\bibitem{Maloy2006}
\bibinfo{author}{M\aa{}l\o{}y, K.~J.}, \bibinfo{author}{Santucci, S.},
  \bibinfo{author}{Schmittbuhl, J.} \& \bibinfo{author}{Toussaint, R.}
\newblock \bibinfo{journal}{\bibinfo{title}{Local waiting time fluctuations
  along a randomly pinned crack front}}.
\newblock {\emph{\JournalTitle{Phys. Rev. Lett.}}}
  \textbf{\bibinfo{volume}{96}}, \bibinfo{pages}{045501},
  \doiprefix\url{10.1103/PhysRevLett.96.045501} (\bibinfo{year}{2006}).

\bibitem{Santucci_2010}
\bibinfo{author}{Santucci, S.} \emph{et~al.}
\newblock \bibinfo{journal}{\bibinfo{title}{Fracture roughness scaling: A case
  study on planar cracks}}.
\newblock {\emph{\JournalTitle{Europhysics Letters}}}
  \textbf{\bibinfo{volume}{92}}, \bibinfo{pages}{44001},
  \doiprefix\url{10.1209/0295-5075/92/44001} (\bibinfo{year}{2010}).

\bibitem{Tallakstad2011}
\bibinfo{author}{Tallakstad, K.~T.}, \bibinfo{author}{Toussaint, R.},
  \bibinfo{author}{Santucci, S.}, \bibinfo{author}{Schmittbuhl, J.} \&
  \bibinfo{author}{Måløy, K.~J.}
\newblock \bibinfo{journal}{\bibinfo{title}{Local dynamics of a randomly pinned
  crack front during creep and forced propagation: An experimental study}}.
\newblock {\emph{\JournalTitle{Phys. Rev. E}}} \textbf{\bibinfo{volume}{83}},
  \bibinfo{pages}{046108}, \doiprefix\url{10.1103/PhysRevE.83.046108}
  (\bibinfo{year}{2011}).

\bibitem{Tanguy1998}
\bibinfo{author}{Tanguy, A.}, \bibinfo{author}{Gounelle, M.} \&
  \bibinfo{author}{Roux, S.}
\newblock \bibinfo{journal}{\bibinfo{title}{From individual to collective
  pinning: Effect of long-range elastic interactions}}.
\newblock {\emph{\JournalTitle{Phys. Rev. E}}} \textbf{\bibinfo{volume}{58}},
  \bibinfo{pages}{1577--1590}, \doiprefix\url{10.1103/PhysRevE.58.1577}
  (\bibinfo{year}{1998}).

\bibitem{a_numerical_study}
\bibinfo{author}{Gjerden, K.~S.}, \bibinfo{author}{Stormo, A.} \&
  \bibinfo{author}{Hansen, A.}
\newblock \bibinfo{journal}{\bibinfo{title}{Local dynamics of a randomly pinned
  crack front: a numerical study}}.
\newblock {\emph{\JournalTitle{Frontiers in Physics}}}
  \textbf{\bibinfo{volume}{2}}, \bibinfo{pages}{66},
  \doiprefix\url{10.3389/fphy.2014.00066} (\bibinfo{year}{2014}).

\bibitem{bundle2}
\bibinfo{author}{Stormo, A.}, \bibinfo{author}{Lengliné, O.},
  \bibinfo{author}{Schmittbuhl, J.} \& \bibinfo{author}{Hansen, A.}
\newblock \bibinfo{journal}{\bibinfo{title}{Soft-clamp fiber bundle model and
  interfacial crack propagation: Comparison using a non-linear imposed
  displacement}}.
\newblock {\emph{\JournalTitle{Frontiers in Physics}}}
  \textbf{\bibinfo{volume}{4}}, \bibinfo{pages}{18},
  \doiprefix\url{10.3389/fphy.2016.00018} (\bibinfo{year}{2016}).

\bibitem{Brenner}
\bibinfo{author}{Brenner, S.~S.}
\newblock \bibinfo{journal}{\bibinfo{title}{Mechanical behavior of sapphire
  whiskers at elevated temperatures}}.
\newblock {\emph{\JournalTitle{Journal of Applied Physics}}}
  \textbf{\bibinfo{volume}{33}}, \bibinfo{pages}{33--39},
  \doiprefix\url{10.1063/1.1728523} (\bibinfo{year}{1962}).

\bibitem{Zhurkov1984}
\bibinfo{author}{Zhurkov, S.~N.}
\newblock \bibinfo{journal}{\bibinfo{title}{Kinetic concept of the strength of
  solids}}.
\newblock {\emph{\JournalTitle{International Journal of Fracture}}}
  \textbf{\bibinfo{volume}{26}}, \bibinfo{pages}{295--307},
  \doiprefix\url{10.1007/BF00962961} (\bibinfo{year}{1984}).

\bibitem{Santucci2004}
\bibinfo{author}{Santucci, S.}, \bibinfo{author}{Vanel, L.} \&
  \bibinfo{author}{Ciliberto, S.}
\newblock \bibinfo{journal}{\bibinfo{title}{Subcritical statistics in rupture
  of fibrous materials: Experiments and model}}.
\newblock {\emph{\JournalTitle{Phys. Rev. Lett.}}}
  \textbf{\bibinfo{volume}{93}}, \bibinfo{pages}{095505},
  \doiprefix\url{10.1103/PhysRevLett.93.095505} (\bibinfo{year}{2004}).

\bibitem{Santucci_2006}
\bibinfo{author}{Santucci, S.}, \bibinfo{author}{Cortet, P.-P.},
  \bibinfo{author}{Deschanel, S.}, \bibinfo{author}{Vanel, L.} \&
  \bibinfo{author}{Ciliberto, S.}
\newblock \bibinfo{journal}{\bibinfo{title}{Subcritical crack growth in fibrous
  materials}}.
\newblock {\emph{\JournalTitle{Europhysics Letters}}}
  \textbf{\bibinfo{volume}{74}}, \bibinfo{pages}{595--601},
  \doiprefix\url{10.1209/epl/i2005-10575-2} (\bibinfo{year}{2006}).

\bibitem{Vanel_2009}
\bibinfo{author}{Vanel, L.}, \bibinfo{author}{Ciliberto, S.},
  \bibinfo{author}{Cortet, P.-P.} \& \bibinfo{author}{Santucci, S.}
\newblock \bibinfo{journal}{\bibinfo{title}{Time-dependent rupture and slow
  crack growth: elastic and viscoplastic dynamics}}.
\newblock {\emph{\JournalTitle{Journal of Physics D: Applied Physics}}}
  \textbf{\bibinfo{volume}{42}}, \bibinfo{pages}{214007},
  \doiprefix\url{10.1088/0022-3727/42/21/214007} (\bibinfo{year}{2009}).

\bibitem{Lengline2011}
\bibinfo{author}{Lengliné, O.} \emph{et~al.}
\newblock \bibinfo{journal}{\bibinfo{title}{Average crack-front velocity during
  subcritical fracture propagation in a heterogeneous medium}}.
\newblock {\emph{\JournalTitle{Phys. Rev. E}}} \textbf{\bibinfo{volume}{84}},
  \bibinfo{pages}{036104}, \doiprefix\url{10.1103/PhysRevE.84.036104}
  (\bibinfo{year}{2011}).

\bibitem{NonGaussian_Fracture}
\bibinfo{author}{Tallakstad, K.}, \bibinfo{author}{Toussaint, R.},
  \bibinfo{author}{Santucci, S.} \& \bibinfo{author}{M\aa{}l\o{}y, K.}
\newblock \bibinfo{journal}{\bibinfo{title}{Non-gaussian nature of fracture and
  the survival of fat-tail exponents}}.
\newblock {\emph{\JournalTitle{Phys. Rev. Lett.}}}
  \textbf{\bibinfo{volume}{110}}, \bibinfo{pages}{145501},
  \doiprefix\url{10.1103/PhysRevLett.110.145501} (\bibinfo{year}{2013}).

\bibitem{TVD1}
\bibinfo{author}{Vincent-Dospital, T.}, \bibinfo{author}{Toussaint, R.},
  \bibinfo{author}{Cochard, A.}, \bibinfo{author}{Måløy, K.~J.} \&
  \bibinfo{author}{Flekkøy, E.~G.}
\newblock \bibinfo{journal}{\bibinfo{title}{Thermal weakening of cracks and
  brittle-ductile transition of matter: {A} phase model}}.
\newblock {\emph{\JournalTitle{Physical Review Materials}}}
  \doiprefix\url{10.1103/PhysRevMaterials.4.023604} (\bibinfo{year}{2020}).

\bibitem{TVD2}
\bibinfo{author}{Vincent-Dospital, T.} \emph{et~al.}
\newblock \bibinfo{journal}{\bibinfo{title}{How heat controls fracture: the
  thermodynamics of creeping and avalanching cracks}}.
\newblock {\emph{\JournalTitle{Soft Matter}}} \doiprefix\url{10.1039/d0sm010}
  (\bibinfo{year}{2020}).

\bibitem{kinetics}
\bibinfo{author}{Hammes, G.~G.}
\newblock \emph{\bibinfo{title}{Principles of Chemical Kinetics}}
  (\bibinfo{publisher}{Academic Press}, \bibinfo{year}{1978}).

\bibitem{Freund1972}
\bibinfo{author}{Freund, L.~B.}
\newblock \bibinfo{journal}{\bibinfo{title}{Crack propagation in an elastic
  solid subjected to general loading}}.
\newblock {\emph{\JournalTitle{Journal of the Mechanics and Physics of
  Solids}}} \textbf{\bibinfo{volume}{20}}, \bibinfo{pages}{129 -- 152},
  \doiprefix\url{10.1016/0022-5096(72)90006-3} (\bibinfo{year}{1972}).

\bibitem{interf_imaging}
\bibinfo{author}{Santucci, S.}, \bibinfo{author}{Måløy, K.~J.},
  \bibinfo{author}{Toussaint, R.} \& \bibinfo{author}{Schmittbuhl, J.}
\newblock \bibinfo{title}{Self-affine scaling during interfacial crack front
  propagation}.
\newblock In \emph{\bibinfo{booktitle}{Dynamics of Complex Interconnected
  Systems}} (\bibinfo{publisher}{NATO ASI, Geilo, Springer},
  \bibinfo{year}{2006}).

\bibitem{PMA3ss}
\bibinfo{author}{Hattali, M.}, \bibinfo{author}{Bar{\'{e}}s, J.},
  \bibinfo{author}{Ponson, L.} \& \bibinfo{author}{Bonamy, D.}
\newblock \bibinfo{title}{Low velocity surface fracture patterns in brittle
  material: A newly evidenced mechanical instability}.
\newblock In \emph{\bibinfo{booktitle}{THERMEC 2011}}, vol.
  \bibinfo{volume}{706} of \emph{\bibinfo{series}{Materials Science Forum}},
  \bibinfo{pages}{920--924},
  \doiprefix\url{10.4028/www.scientific.net/MSF.706-709.920}
  (\bibinfo{publisher}{Trans Tech Publications Ltd}, \bibinfo{year}{2012}).

\bibitem{Perrin_spectral_redistribution}
\bibinfo{author}{Perrin, G.}, \bibinfo{author}{Rice, J.~R.} \&
  \bibinfo{author}{Zheng, G.}
\newblock \bibinfo{journal}{\bibinfo{title}{Self-healing slip pulse on a
  frictional surface}}.
\newblock {\emph{\JournalTitle{Journal of the Mechanics and Physics of
  Solids}}} \textbf{\bibinfo{volume}{43}}, \bibinfo{pages}{1461 -- 1495},
  \doiprefix\url{10.1016/0022-5096(95)00036-I} (\bibinfo{year}{1995}).

\bibitem{DormandPrince}
\bibinfo{author}{Dormand, J.~R.} \& \bibinfo{author}{Prince, P.~J.}
\newblock \bibinfo{journal}{\bibinfo{title}{A family of embedded
  {R}unge-{K}utta formulae}}.
\newblock {\emph{\JournalTitle{Journal of Computational and Applied
  Mathematics}}} \textbf{\bibinfo{volume}{6}}, \bibinfo{pages}{19 -- 26},
  \doiprefix\url{10.1016/0771-050X(80)90013-3} (\bibinfo{year}{1980}).

\bibitem{dop583}
\bibinfo{author}{Hairer, E.}, \bibinfo{author}{Nørsett, S.~P.} \&
  \bibinfo{author}{Wanner, G.}
\newblock \emph{\bibinfo{title}{Solving Ordinary Differential Equations I,
  nonstiff problems}} (\bibinfo{publisher}{Springer-Verlag Berlin Heidelberg},
  \bibinfo{year}{1993}).

\bibitem{plexi}
\bibinfo{author}{Zerwer, A.}, \bibinfo{author}{Polak, M.~A.} \&
  \bibinfo{author}{Santamarina, J.~C.}
\newblock \bibinfo{journal}{\bibinfo{title}{Wave propagation in thin plexiglas
  plates: implications for {R}ayleigh waves}}.
\newblock {\emph{\JournalTitle{NDT and E International}}}
  \textbf{\bibinfo{volume}{33}}, \bibinfo{pages}{33 -- 41},
  \doiprefix\url{10.1016/S0963-8695(99)00010-9} (\bibinfo{year}{2000}).

\bibitem{fractal_surface}
\bibinfo{author}{Barabási, A.-L.} \& \bibinfo{author}{Stanley, H.~E.}
\newblock \emph{\bibinfo{title}{Fractal Concepts in Surface Growth}}
  (\bibinfo{publisher}{Cambridge University Press}, \bibinfo{year}{1995}).

\bibitem{cor_low_t}
\bibinfo{author}{Soriano, J.} \emph{et~al.}
\newblock \bibinfo{journal}{\bibinfo{title}{Anomalous roughening of viscous
  fluid fronts in spontaneous imbibition}}.
\newblock {\emph{\JournalTitle{Phys. Rev. Lett.}}}
  \textbf{\bibinfo{volume}{95}}, \bibinfo{pages}{104501},
  \doiprefix\url{10.1103/PhysRevLett.95.104501} (\bibinfo{year}{2005}).

\bibitem{confpaper}
\bibinfo{author}{Måløy, K.~J.}, \bibinfo{author}{Toussaint, R.} \&
  \bibinfo{author}{Schmittbuhl, J.}
\newblock \bibinfo{title}{Dynamics and structure of interfacial crack front}.
\newblock In \emph{\bibinfo{booktitle}{11th International Conference on
  Fracture 2005, ICF11}}, \bibinfo{number}{7} (\bibinfo{year}{2005}).

\bibitem{altuglas}
\bibinfo{title}{Technical information, {Altuglas sheets}}.
\newblock \bibinfo{type}{Tech. Rep.}, \bibinfo{institution}{Arkema}
  (\bibinfo{year}{2017}).

\bibitem{Marshall_1974}
\bibinfo{author}{Marshall, G.~P.}, \bibinfo{author}{Coutts, L.~H.} \&
  \bibinfo{author}{Williams, J.~G.}
\newblock \bibinfo{journal}{\bibinfo{title}{Temperature effects in the fracture
  of {PMMA}}}.
\newblock {\emph{\JournalTitle{Journal of Materials Science}}}
  \textbf{\bibinfo{volume}{9}}, \bibinfo{pages}{1409--1419},
  \doiprefix\url{10.1007/BF00552926} (\bibinfo{year}{1974}).

\bibitem{Santucci_2018}
\bibinfo{author}{Santucci, S.} \emph{et~al.}
\newblock \bibinfo{journal}{\bibinfo{title}{Avalanches and extreme value
  statistics in interfacial crackling dynamics}}.
\newblock {\emph{\JournalTitle{Philosophical Transactions of the Royal Society
  A: Mathematical, Physical and Engineering Sciences}}}
  \textbf{\bibinfo{volume}{377}}, \bibinfo{pages}{20170394},
  \doiprefix\url{10.1098/rsta.2017.0394} (\bibinfo{year}{2019}).

\bibitem{bouchaud_coales}
\bibinfo{author}{Bouchaud, E.}, \bibinfo{author}{Bouchaud, J.},
  \bibinfo{author}{Fisher, D.}, \bibinfo{author}{Ramanathan, S.} \&
  \bibinfo{author}{Rice, J.}
\newblock \bibinfo{journal}{\bibinfo{title}{Can crack front waves explain the
  roughness of cracks?}}
\newblock {\emph{\JournalTitle{Journal of the Mechanics and Physics of
  Solids}}} \textbf{\bibinfo{volume}{50}}, \bibinfo{pages}{1703 -- 1725},
  \doiprefix\url{https://doi.org/10.1016/S0022-5096(01)00137-5}
  (\bibinfo{year}{2002}).

\bibitem{fiber_universel}
\bibinfo{author}{Gjerden, K.~S.}, \bibinfo{author}{Stormo, A.} \&
  \bibinfo{author}{Hansen, A.}
\newblock \bibinfo{journal}{\bibinfo{title}{Universality classes in constrained
  crack growth}}.
\newblock {\emph{\JournalTitle{Phys. Rev. Lett.}}}
  \textbf{\bibinfo{volume}{111}}, \bibinfo{pages}{135502},
  \doiprefix\url{10.1103/PhysRevLett.111.135502} (\bibinfo{year}{2013}).

\bibitem{Scorretti_2001}
\bibinfo{author}{Scorretti, R.}, \bibinfo{author}{Ciliberto, S.} \&
  \bibinfo{author}{Guarino, A.}
\newblock \bibinfo{journal}{\bibinfo{title}{Disorder enhances the effects of
  thermal noise in the fiber bundle model}}.
\newblock {\emph{\JournalTitle{Europhysics Letters}}}
  \textbf{\bibinfo{volume}{55}}, \bibinfo{pages}{626--632},
  \doiprefix\url{10.1209/epl/i2001-00462-x} (\bibinfo{year}{2001}).

\bibitem{roux}
\bibinfo{author}{Roux, S.}
\newblock \bibinfo{journal}{\bibinfo{title}{Thermally activated breakdown in
  the fiber-bundle model}}.
\newblock {\emph{\JournalTitle{Phys. Rev. E}}} \textbf{\bibinfo{volume}{62}},
  \bibinfo{pages}{6164--6169}, \doiprefix\url{10.1103/PhysRevE.62.6164}
  (\bibinfo{year}{2000}).

\bibitem{ToussaintSoft}
\bibinfo{author}{Toussaint, R.} \emph{et~al.}
\newblock \bibinfo{journal}{\bibinfo{title}{How cracks are hot and cool: a
  burning issue for paper}}.
\newblock {\emph{\JournalTitle{Soft Matter}}} \textbf{\bibinfo{volume}{12}},
  \bibinfo{pages}{5563--5571}, \doiprefix\url{10.1039/C6SM00615A}
  (\bibinfo{year}{2016}).

\bibitem{TVD3}
\bibinfo{author}{Vincent-Dospital, T.}, \bibinfo{author}{Toussaint, R.},
  \bibinfo{author}{Cochard, A.}, \bibinfo{author}{Flekkøy, E.~G.} \&
  \bibinfo{author}{Måløy, K.~J.}
\newblock \bibinfo{journal}{\bibinfo{title}{Thermal dissipation as both the
  strength and weakness of matter. a material failure prediction by monitoring
  creep}}.
\newblock {\emph{\JournalTitle{Soft Matter}}}
  \doiprefix\url{10.1039/D0SM02089C} (\bibinfo{year}{2021}).

\bibitem{dugdale1960}
\bibinfo{author}{Dugdale, D.}
\newblock \bibinfo{journal}{\bibinfo{title}{Yielding of steel sheets containing
  slits}}.
\newblock {\emph{\JournalTitle{Journal of the Mechanics and Physics of
  Solids}}} \textbf{\bibinfo{volume}{8}}, \bibinfo{pages}{100 -- 104},
  \doiprefix\url{10.1016/0022-5096(60)90013-2} (\bibinfo{year}{1960}).

\bibitem{Entangled}
\bibinfo{author}{Henkee, C.~S.} \& \bibinfo{author}{Kramer, E.~J.}
\newblock \bibinfo{journal}{\bibinfo{title}{Crazing and shear deformation in
  crosslinked polystyrene}}.
\newblock {\emph{\JournalTitle{Journal of Polymer Science: Polymer Physics
  Edition}}} \textbf{\bibinfo{volume}{22}}, \bibinfo{pages}{721--737},
  \doiprefix\url{10.1002/pol.1984.180220414} (\bibinfo{year}{1984}).

\bibitem{fracmec}
\bibinfo{author}{Anderson, T.~L.}
\newblock \emph{\bibinfo{title}{Fracture Mechanics: Fundamentals and
  Applications}} (\bibinfo{publisher}{Taylor and Francis},
  \bibinfo{year}{2005}).

\end{thebibliography}

\end{document}